\documentclass[twoside,12pt]{article}

\usepackage{epsfig}

\usepackage{graphicx}
\usepackage{amsmath}
\usepackage{amssymb}
\usepackage{array}
\usepackage{wrapfig}
\usepackage{units}
\usepackage{caption}
\usepackage{slashed}
\usepackage{color}

\newcommand{\br}[1]{\:\langle{#1}|\:}
\newcommand{\be}{\begin{equation}}
\newcommand{\ee}{\end{equation}}
\newcommand{\bea}{\begin{eqnarray}}
\newcommand{\eea}{\end{eqnarray}}

\newcommand{\ba}{\begin{eqnarray}}
\newcommand{\ea}{\end{eqnarray}}
\newcommand{\s}{\scriptscriptstyle}

\newcommand{\ket}[1]{\left\lvert #1\right\rangle}

\newcommand{\eS}{{\epsilon_S}}
\newcommand{\eT}{{\epsilon_T}}
\newcommand{\eP}{{\epsilon_P}}

\newcommand{\teS}{{\tilde{\epsilon}_S}}
\newcommand{\teT}{{\tilde{\epsilon}_T}}
\newcommand{\teP}{{\tilde{\epsilon}_P}}
\newcommand{\teL}{{\tilde{\epsilon}_L}}
\newcommand{\teR}{{\tilde{\epsilon}_R}}

\newcommand{\nc}{\newcommand}

\nc{\newsection}[1]{\section{#1}\setcounter{equation}{0}}
\nc{\newappendix}[1]{\section*{#1}\setcounter{equation}{0}}
\nc{\scm}{\scriptscriptstyle\mathrm}
\nc{\f}{\frac}
\nc{\baa}{\begin{array}}      \nc{\eaa}{\end{array}}
\nc{\bit}{\begin{itemize}}    \nc{\eit}{\end{itemize}}
\nc{\ben}{\begin{enumerate}}  \nc{\een}{\end{enumerate}}
\nc{\bce}{\begin{center}}     \nc{\ece}{\end{center}}
\nc{\bfl}{\begin{flushright}} \nc{\efl}{\end{flushright}}
\nc{\btb}{\begin{tabular}}    \nc{\etb}{\end{tabular}}
\nc{\eps}{\varepsilon}
\nc{\vp}{\varphi}
\nc{\tvp}{\widetilde{\varphi}}
\nc{\D}{\mbox{$\not\!\!D$}}
\nc{\Db}{\mbox{${\raisebox{2mm}{\boldmath ${}^\leftarrow$}\hspace{-4mm} D}$}}
\nc{\Dfb}{\mbox{$\raisebox{2mm}{\boldmath ${}^\leftrightarrow$}\hspace{-4mm} D$}}
\nc{\vpj }{\mbox{${\vp^\dag i\,\raisebox{2mm}{\boldmath ${}^\leftrightarrow$}\hspace{-4mm} D_\mu\,\vp}$}}
\nc{\vpjt}{\mbox{${\vp^\dag i\,\raisebox{2mm}{\boldmath ${}^\leftrightarrow$}\hspace{-4mm} D_\mu^{\,a}\,\vp}$}}
\newcommand\slurp[1]{#1}
\newcommand\sslurp[2]{#1{#2}}
{\catcode`/=\active \expandafter}%
\slurp{\newcommand/}{/\penalty1000\hskip0pt\relax}
{\catcode`:=\active \expandafter}%
\slurp{\newcommand:}{:\penalty1000\hskip0pt\relax}

\newcommand\addspace{\ifcat\nextchar a\spacefactor999. \else.\fi}
{\catcode`\.=\active \expandafter}%
\slurp{\newcommand.}{\futurelet\nextchar\addspace}

\usepackage[unicode]{hyperref}
\ifx\href\undefined\def\href#1{}\fi
\ifx\texorpdfstring\undefined\def\texorpdfstring#1#2{#1}\fi

\newcommand\myslash{/} \newcommand\mycolon{:}
\newcommand\doi{{\catcode`/=\active \catcode`:=\active \expandafter}\sslurp\realdoi}
{\catcode`/=\active \catcode`:=\active \expandafter}%
\slurp{\newcommand\realdoi[1]{{\let/=\myslash \let:=\mycolon
                               \edef\raw{{http://dx.doi.org/#1}}\expandafter}%
                               \expandafter\href\raw{doi:#1}}}
\newcommand\eprint[2]{{\escapechar-1%
                       \edef\a{\expandafter\string\csname arXiv\endcsname}%
                       \edef\b{\expandafter\string\csname #1\endcsname}%
                       \edef\c{\expandafter\string\csname #2\endcsname}%
                       \edef\d{\noexpand\href{http://arXiv.org/abs/\c}}%
                       \ifx\a\b\expandafter\d\fi{\tt #1:#2}}}

\newcommand{\lsim}{\buildrel < \over {_\sim}}
\newcommand{\gsim}{\buildrel > \over {_\sim}}

\begin{document}

\title{ \vspace{1cm}
Beta Decays and Non-Standard Interactions in the LHC Era
}
\author{
Vincenzo Cirigliano$^{1}$, 
Susan Gardner$^2$, and
 Barry Holstein$^3$
\\
\\
$^1$ Theoretical Division, Los Alamos National Laboratory, \\ Los Alamos, NM 87545, USA
\\
$^2$  Department of Physics and Astronomy, University of Kentucky, \\  Lexington, KY 40506-0055 USA
\\
$^3$
Department of Physics-LGRT, University of Massachusetts,\\
Amherst, MA 01003 USA}
\maketitle
\begin{abstract}

We consider the role of precision measurements
of beta decays and light meson semi-leptonic decays    
in probing physics beyond the Standard Model in the LHC era. 
We describe all  low-energy charged-current processes  within and beyond the
Standard Model using  an effective field  theory framework.
We first  discuss the theoretical hadronic input which in these precision tests 
plays  a crucial role in setting the baseline for new physics  searches.
We then review the current and upcoming  constraints on the various non-standard operators 
from the study of decay rates, spectra, and correlations in a broad array of light-quark systems. 
We finally discuss the interplay with LHC searches, both within models and in an  effective theory approach.
Our discussion  illustrates the independent yet complementary nature of 
precision beta decay measurements as probes of new physics, 
showing them to be of continuing importance throughout the LHC era.

\end{abstract}

\newpage

\section{Introduction}

Beta decays played a central role
in determining the $V-A$ structure of the weak interactions and in shaping 
what we now call the Standard Model (SM)~\cite{Feynman:1958ty,Sudarshan:1958vf,Weinberg:2009zz}.
We focus here on the set of semi-leptonic ``charged current" (CC) processes that are
mediated in the SM by tree-level W exchange, up to radiative corrections.
In the SM the CC weak processes are characterized by two main  features:
(i) the hadronic and leptonic bilinear densities involved in the process
have a dominant $V-A$ component, with 
other types of couplings---$V+A,S,P,T$---arising at higher order in radiative corrections
or in recoil momentum;
(ii) the effective Fermi constants extracted in beta decays  obey
lepton universality as well as 
quark-lepton, or Cabibbo, universality,
which is  equivalent in the SM to the unitarity of the Cabibbo-Kobayashi-Maskawa (CKM)  quark
mixing matrix. 
Universality relations can only emerge  once the process-dependent radiative corrections are removed.
Currently precision beta-decay  measurements
involving neutrons, nuclei, and mesons
are used to probe the existence of non-SM interactions which effectively induce violations
of the universality relations  and/or  novel non-$(V-A)$ structures
or corrections to the dominant vector and axial-vector couplings.~\footnote{In 
this review we consider the decays involving the light quarks $u,d$, and $s$ exclusively.}  
The low-energy charged-current-interaction Hamiltonian 
is sensitive to many classes of SM extensions. 
In this sense,  beta decay measurements  can be considered as  ``broad band"  probes of
physics beyond the Standard Model (BSM):
while by themselves they do not allow us  to reconstruct the ultraviolet dynamics,
 they provide, at 0.1\%-level  precision, powerful boundary conditions and diagnostics on virtually any TeV-scale SM extension.

Considerable experimental progress is ongoing or  expected in a few-year time scale on several fronts,
using both cold and ultracold
neutrons~\cite{Abele:2008zz,Dubbers:2011ns,RevModPhys.83.1173,
Dewey2009189,Arzumanov2009186,Walstrom200982,Materne2009176,Leung2009181,PERKEOIII:2009,Plaster:2008si,abBA,Dubbers:2007st,WilburnUCNB,Pocanic:2008pu,aSPECT:2008,Wietfeldt:2005wz,UCNb}, 
trapped nuclei~\cite{Severijns:2011zz,Knecht201143}, and rare pion
and kaon decays~\cite{Pocanic:2012zz,Malbrunot:2011zz,NA62:2011aa,Bryman:2011zz}. 
Some of the measurements plan to reach sensitivities between  $10^{-3}$ and $10^{-4}$;  this makes such
observables  very interesting  probes of new physics effects originating at  the TeV scale, 
because such effects are expected to have size
${\cal O}((v/\Lambda_{\rm BSM})^2)$, where $v = (2 \sqrt{2} G_F)^{-1/2} \approx 174$ GeV
and $\Lambda_{\rm BSM}$ denotes the mass scale where  BSM physics appears.

As in  previous reviews~\cite{Herczeg2001vk,Erler:2004cx,Severijns2006dr,RamseyMusolf:2006vr},
the  overall goal of this  article  is to discuss    the  discovery potential  and discriminating power
of planned precision beta-decay  measurements  with  neutrons, nuclei, and mesons,
in light of other existing precision electroweak tests and  high-energy collider searches, such 
as at the Tevatron and the LHC. 
In order to achieve our goal,  we work 
within an effective field theory (EFT) framework, in which the dynamical
effects of new heavy BSM  degrees of freedom are  parameterized by local operators 
of dimension higher than four
built with SM fields.~\footnote{The EFT analysis can be applied to all low-energy probes of
CC interactions.  It is also valid for collider searches as long as
the particles which mediate the new interactions are above particle-production 
threshold at the operating center-of-mass energy. 
In this case, a direct comparison of low-energy 
and collider constraints can be performed, as we discuss in Section~\ref{sect:collider}.}
All model-specific analyses of beta decays 
can be cast in the EFT language and the 
limits on the effective operators we derive can be readily
converted into constraints on the parameters of any SM extension.
In the absence of an emerging  picture of new dynamics 
from collider searches, the EFT analysis 
is the first necessary step to establish the  motivation and significance
of this set of low-energy probes.
Subsequently, we will also discuss 
well-motivated  models such as the Left-Right Symmetric Model 
and supersymmetric extensions of the SM in order to show the
discriminating power that combinations of beta decay measurements can have on explicit models.

Probing  short-distance  BSM couplings through precision phenomenology of beta decays requires
knowing the relevant hadronic and nuclear  matrix elements to a precision comparable to the
size of the new physics effects one could expect to appear. 
This means that 
one needs to know the hadronic matrix elements of the SM operators, that is, of the 
V and A currents, to the level of 
${\cal O}((v/\Lambda_{\rm BSM})^2$), i.e. of $10^{-3}$ or better. 
This is a necessary condition for beta decays to function as competitive  probes:
we are in search of a small BSM signal, 
and hence we need to know the SM ``background"  to a level comparable to
that of the signal for which we are looking. 
One also needs to know  the matrix elements of the BSM operators, such as the S, T, P densities,
because all the observables are sensitive to the product of
the short-distance BSM coupling with the appropriate hadronic/nuclear matrix element.
Consequently 
if a certain matrix element is suppressed, the sensitivity to the corresponding BSM coupling
is also suppressed. 
Moreover, were such an 
anomalous suppression absent, 
the fractional uncertainty on the BSM matrix element still determines how well
we can constrain that BSM coupling. 
For BSM operators, the precision required of the relevant hadronic matrix element is
much less severe; an uncertainty at the ${\cal O}(10\%)$ level is acceptable.
Motivated by these considerations,
we pay special attention to the hadronic and nuclear uncertainties which appear.

This paper is organized as follows. 
In Section~\ref{sect:framework}  we set up the theoretical framework
for the analysis of all low-energy CC processes within and beyond the SM. 
In Section  \ref{sect:universality} we discuss the status  of Cabibbo universality tests
(Sec.~\ref{sect:CKM}) and lepton universality tests
(Sec.~\ref{sect:LFU}) and explore the implications for BSM physics.
In Section~\ref{sect:correlations} we focus on differential decay distributions in beta decays
and discuss the implications for non-$(V-A)$ couplings.
In Section \ref{sect:collider} we explore the constraints on non-standard CC couplings that can
be obtained from LHC data.
In Section  \ref{sect:models}
we illustrate how the precision tests can be used to probe the parameter space of  models
such as the Left-Right Symmetric Model and the 
Minimal Supersymmetric Standard Model (MSSM),
and we present our concluding remarks in
 Section \ref{sect:conclusions}.

\section{Theoretical Framework \label{sect:framework}}

\subsection{\it  Effective Lagrangian}

In this review we take the point of view that the Standard Model emerges as the low-energy limit of a more fundamental   
theory  characterized by the scale 
$\Lambda$ at which new particles appear. Consequently, at energies scales below $\Lambda$,
namely, $\Lambda > E >   M_{Z,W}$, 
the new degrees of freedom are no longer present; they have been 
``integrated out," yielding an effective Lagrangian comprised of 
the SM Lagrangian augmented by a string of
$d>4$  operators constructed with the low-energy SM fields, suppressed by 
$\Lambda^{d-4}$~\cite{Appelquist:1974tg}, 
that respect the  SU(3)$_C\times$SU(2)$_L\times$U(1)$_Y$  gauge symmetry of the SM. 
Flavor physics observables constrain the appearance of non-SM invariant operators to energies
far beyond the weak scale~\cite{Isidori:2010kg,Bona:2007vi,Charles:2004jd}. 
The building blocks of the gauge-invariant  local operators are: 
the  gauge fields $G_\mu^A,  \, W_\mu^a, \, B_\mu$, 
corresponding to SU(3)$_C\times$ SU(2)$_L \times$ U(1)$_Y$, respectively, 
the six fermionic gauge multiplets, including a singlet right-handed neutrino state, 
\begin{equation}
q^i =
\left(
\begin{array}{c}
u_L^i \\
d_L^i
\end{array}
\right)  \qquad
u^i = u_R^i \qquad
d^i = d_R^i \qquad 
l^i =
\left(
\begin{array}{c}
\nu_L^i\\
e_L^i
\end{array}
\right)
\qquad e^i = e_R^i
\qquad \nu^i = \nu_R^i~,  
\label{eq:fermions}
\end{equation}
the Higgs doublet $\varphi$
\begin{equation}
\varphi =
\left(
\begin{array}{c}
\varphi^+ \\
\varphi^0
\end{array}
\right)~,
\end{equation}
and the covariant derivative
\begin{equation}
D_{\mu} =   I \, \partial_\mu   \, - \, i g_s \frac{\lambda^A}{2} G_\mu^A \,
- \, i g \frac{\sigma^a}{2} W_\mu^a   \, - \, i g'  Y B_\mu~.
\end{equation}
In the above expression, $I$ is the identity matrix, 
$\lambda^A$ are the SU(3) Gell-Mann matrices; 
$\sigma^a$ are the SU(2)  Pauli matrices;  $g_s, g$, and $g'$ 
are the gauge couplings; and $Y$ is the hypercharge of a given multiplet. The introduction 
of three light, right-handed neutrinos to accommodate the existence of neutrino 
masses illustrates
explicitly  that  new light degrees of freedom can be included in the EFT 
if  we assign their SU(3)$_C\times$SU(2)$_L\times$U(1)$_Y$  quantum numbers. 
The impact of other light, new physics is left as an exercise for future work.

The leading operators which modify CC interactions are of dimension six, 
though it is worth noting
that new physics can also modify the radiative corrections and hence 
the couplings with which the SM operators appear.  
The minimal set of operators contributing to low-energy semi-leptonic charged current processes can 
be divided into two groups: operators built out of SM fields, noting the left column below, 
in which we follow the notation of Refs.~\cite{Buchmuller:1985jz,Grzadkowski:2010es}, 
and  operators involving the singlet  R-handed neutrino field  $\nu$~\cite{Cirigliano:2012ab}, which are 
displayed in the right column below.
Furthermore, within each group  the operators can be divided into two classes---four-fermion 
contact interactions and 
vertex corrections. The vertex correction operators are written in SU(2)-invariant form and
therefore involve the Higgs doublet: after electroweak symmetry breaking (EWSB) 
they include terms involving a $W$ (or $Z$) boson, 
a fermion, and an anti-fermion. 
Here is the list: 

\noindent{\underline {Four-fermion operators}}:
\begin{subequations}
\label{eq:ws1}
\bea
&\!\!\!\!\!\!\!\!\!
O_{l q}^{(3)}= (\overline{l} \gamma^\mu \sigma^a l) (\overline{q} \gamma_\mu \sigma^a q)~~~~~~~~~~~~~~~~~~~
&O_{e\nu ud} = (\overline{e} \gamma^\mu \nu) (\overline{u} \gamma_\mu  d)
+ {\rm h.c.}~~~~~~~~
\\
&\!\!\!\!\!\!
O_{ledq} = (\overline{l} e) (\overline{d} q)+ {\rm h.c.}~~~~~~~~~~~~~~~~~~~~~~~
&O_{qu\nu} = (\overline{l} \nu) (\overline{u} q)+ {\rm h.c.}~~~~~~~~~~
\\
&
O_{l e q u}^{(1)} = (\bar{l}_a e)\epsilon^{ab}(\bar{q}_b u)+ {\rm h.c.}~~~~~~~~~~~~~~~~~~~~~
&O_{l \nu  q d}^{(1)} = (\bar{l}_a \nu)\epsilon^{ab}(\bar{q}_b d)+ {\rm h.c.}~~~~~
\\
&
O_{le qu}^{(3)} =(\bar{l}_a\sigma^{\mu\nu}e)\epsilon^{ab}(\bar{q}_b\sigma_{\mu\nu}u)+ {\rm h.c.}~~~~~~~~~~~~
&O_{l \nu  qd}^{(3)} =(\bar{l}_a\sigma^{\mu\nu} \nu)\epsilon^{ab}(\bar{q}_b\sigma_{\mu\nu}d)+ {\rm h.c.}~~~~~~~~~~
\eea
\end{subequations}

\noindent{\underline{Vertex corrections}:
\begin{subequations}
\label{eq:ws2}
\bea
&& O_{\varphi u d} = i(\varphi^T \epsilon  D_\mu \varphi) (\overline{u}\gamma^\mu d)+ {\rm h.c.}~~~~~~~~~
      O_{\varphi \nu e} ' = i(\varphi^T \epsilon  D_\mu \varphi) (\overline{\nu}\gamma^\mu e)+ {\rm h.c.}~~~~~~~~~
\\
&&
O_{\varphi q}^{(3)} = \!
(\vpjt)
 (\overline{q} \gamma_\mu \sigma^a q)
\\
&&
O_{\varphi l}^{(3)} = \!
(\vpjt)
(\overline{l} \gamma_\mu \sigma^a l)~. 
\eea
\end{subequations}

Denoting by $\Lambda_i$  the effective dimensionful coupling associated with the operator $O_i$,
we can write the effective Lagrangian as
\bea
{\cal L}^{(\rm{eff})}
= {\cal L}_{\rm{SM}} + \sum_{i} \frac{1}{\Lambda_i^2}~ O_i \ \longrightarrow \
{\cal L}_{\rm{SM}} +  \frac{1}{v^2}  \, \sum_{i}  \, \hat{\alpha}_i   ~ O_i \, ,
\qquad
{\rm with}  \  \  \hat{\alpha}_i = \frac{v^2}{\Lambda_i^2}~,
\eea
where in the last step we have set the correct dimensions using the
Higgs vacuum expectation value (VEV) $v = \langle \varphi^0 \rangle =   (2 \sqrt{2} G_F)^{-1/2}$
and defined the dimensionless  new-physics couplings $\hat{\alpha}_i$, 
which in general are matrices in both the quark and lepton flavor spaces.
In this framework one can derive the low-energy effective Lagrangian at 
${\cal O}(1 \ {\rm GeV})$  for semi-leptonic transitions. 
It  receives contributions from both $W$-exchange diagrams, with modified $W$-fermion couplings, 
and the four-fermion operators.
After including the electroweak radiative corrections to the SM operator~\cite{Sirlin:1981ie}, 
the matching procedure
leads to a low-energy quark level  effective Lagrangian involving ten dimension-six operators:
\bea
{\cal L}_{\rm CC}  &=&
- \frac{G_F^{(0)} V_{ud}}{\sqrt{2}} \  \Big[ 
\ \Big( 1   + \delta_{\beta} \Big) \
\bar{e}  \gamma_\mu  (1 - \gamma_5)   \nu_{e}  \cdot \bar{u}   \gamma^\mu  (1 - \gamma_5)  d
\label{eq:leff10}
 \\
& + &
\ \epsilon_L  \
\bar{e}  \gamma_\mu  (1 - \gamma_5)   \nu_{\ell}  \cdot \bar{u}   \gamma^\mu  (1 - \gamma_5)  d
+
\tilde{\epsilon}_L  \ \ \bar{e}  \gamma_\mu  (1 + \gamma_5)   \nu_{\ell}  \cdot \bar{u}   \gamma^\mu  (1 - \gamma_5)  d
\nonumber\\
&+&   \epsilon_R   \  \   \bar{e}  \gamma_\mu  (1 - \gamma_5)   \nu_{\ell}
\cdot \bar{u}   \gamma^\mu  (1 + \gamma_5)  d
\  + \
\tilde{ \epsilon}_R   \  \   \bar{e}  \gamma_\mu  (1 +  \gamma_5)   \nu_{\ell}
\cdot \bar{u}   \gamma^\mu  (1 + \gamma_5)  d
\nonumber\\
&+&  \epsilon_S  \  \  \bar{e}  (1 - \gamma_5) \nu_{\ell}  \cdot  \bar{u} d
 \ + \  \tilde{\epsilon}_S  \  \  \bar{e}  (1 +  \gamma_5) \nu_{\ell}  \cdot  \bar{u} d
 \nonumber \\
 &-& \epsilon_P  \  \   \bar{e}  (1 - \gamma_5) \nu_{\ell}  \cdot  \bar{u} \gamma_5 d
 \ -  \  \tilde{\epsilon}_P  \  \   \bar{e}  (1 + \gamma_5) \nu_{\ell}  \cdot  \bar{u} \gamma_5 d
 \nonumber \\
 &+&
\epsilon_T    \   \bar{e}   \sigma_{\mu \nu} (1 - \gamma_5) \nu_{\ell}    \cdot  \bar{u}   \sigma^{\mu \nu} (1 - \gamma_5) d
\ + \
\tilde{\epsilon}_T      \   \bar{e}   \sigma_{\mu \nu} (1 + \gamma_5) \nu_{\ell}    \cdot  \bar{u}
 \sigma^{\mu \nu} (1 + \gamma_5) d + {\rm h.c.}~.
\nonumber
\eea
In the above equation  $G_F^{(0)}/\sqrt{2} =  g^2/(8 M_W^2)$ is the tree-level SM Fermi constant, 
and $\delta_{\beta}$ encodes the effect of SM electroweak radiative 
corrections to semi-leptonic transitions, noting that the Fermi theory QED contributions have been 
subtracted~\cite{Sirlin:1977sv,Sirlin:1974ni,Sirlin:1981ie,Marciano:1985pd,Czarnecki:2004cw,Marciano:2005ec}. 
The coupling $G_F^{(0)}$ can be expressed in terms of the Fermi constant   $G_\mu = 1.166371 (6) \times 10^{-5} {\rm GeV}^{-2}$ 
precisely measured in muon decay~\cite{Chitwood:2007pa}.  
In order to do so, one has to consider the  low-energy effective Lagrangian 
describing muon decay~\cite{Cirigliano:2009wk}, 
\be
{\cal L}_{\mu \to e \bar{\nu}_e \nu_\mu}
=  - 4 \,  G_F^{(0)}
\left(1 + \delta_{\mu} + \epsilon_\mu
\right) \ 
\bar{e}_{\s{L}} \gamma_\mu \nu_{e\s{L}} \cdot 
 \bar{\nu}_{\mu \s{L}} \gamma^\mu \mu_{\s{L}}
~+~ \rm{ h.c.}~, 
\label{eq:leffmu}
\ee
where $G_\mu \equiv G_F^{(0)} (1 + \delta_\mu + \epsilon_\mu) $. 
Here $\delta_{\mu}$ represents  the SM electroweak 
radiative corrections~\cite{vanRitbergen:1999fi} to purely leptonic transitions, 
noting that the Fermi theory QED contributions have been subtracted, 
and $\epsilon_\mu$ encodes possible new physics contributions, 
so that $G_F^{(0)} = G_\mu ( 1 - \delta_{\mu} - \epsilon_\mu)$~\footnote{Our notation in Eqs.(\ref{eq:leff10}) and
(\ref{eq:leffmu}) corresponds to that of Ref.~\cite{RamseyMusolf:2006vr} if we 
replace  $\delta_\beta  \to \Delta \hat{r}_\beta$ 
and $\delta_\mu  \to \Delta \hat{r}_\mu$.}. 

The BSM effective couplings in Eq.~(\ref{eq:leff10})  are denoted by $\epsilon_\alpha$ and  $\tilde{\epsilon}_\beta$,
using the self-explanatory 
notation  $\alpha, \beta = L,R,S,P,T$. These couplings 
can be expressed in terms of
the weak-scale couplings $\hat{\alpha}_j$~\cite{Cirigliano:2009wk, Bhattacharya:2011qm,Cirigliano:2012ab}.
In the 
effective Lagrangian of Eq.~(\ref{eq:leff10}), 
$e,u$, and $d$ denote the electron, up-, and down-quark
mass eigenfields, whereas $\nu_\ell$ represents the neutrino
flavor fields. In general we can have $\ell \neq e$---in what follows, 
we suppress lepton flavor indices. 
Finally,  identical  CC  effective operators appear for other  quark flavors.
For example,  the operators obtained by replacing  the  $d$ quark  with the strange quark $s$
describe $|\Delta S| = 1$ semileptonic processes.

Next, we discuss some  noteworthy  points in regards to 
the effective Lagrangian of Eq.~(\ref{eq:leff10}):

\begin{itemize} 
\item The effective couplings denoted by $\epsilon_\alpha$ involve L-handed neutrinos,
whereas  $\tilde{\epsilon}_\beta$ involve R-handed neutrinos.  Therefore,
the $\tilde{\epsilon}_\beta$ appear in decay rates and distributions  either quadratically or
linearly, but the latter appears 
multiplied by the small factor $m_\nu/E_\nu$, as it is realized 
through interference of the SM and BSM couplings.
In constrast, the $\epsilon_\alpha$ couplings contribute linearly to the decay rates
without  $m_\nu/E_\nu$ suppression.
As a consequence, the bounds on the $\epsilon$'s are much stronger than the bounds
on the $\tilde{\epsilon}$'s.

\item
There are twelve SU(2)${}_L\times$U(1)-invariant 
operators that contribute to beta decays, though
 there are only ten quark-level U(1)$_{EM}$-invariant operators.
This is because the correction $\epsilon_L$ to the SM operator
encodes contributions from three weak-scale operators of 
Eqs.~(\ref{eq:ws1}) and (\ref{eq:ws2}), 
namely, the contact operator  $O_{lq}^{(3)}$ and the quark and lepton vertex corrections,
$O_{\varphi q}^{(3)}$ and  $O_{\varphi l}^{(3)}$. All other low-energy operators are in 
one-to-one correspondence with the TeV scale SU(2)${}_L\times$U(1)-invariant operators. 
It is interesting to note that SU(2) gauge invariance implies that 
the same  couplings mediate not only charged-current processes but also  ``neutral current"
processes such as $\bar{e} e \leftrightarrow \bar{u} u, \bar{d} d$.

\item While the physical amplitudes  are  renormalization scale  and scheme independent, 
the individual effective couplings 
$\epsilon_{S,P,T}$  ($\tilde{\epsilon}_{S,P,T}$)   and the corresponding hadronic matrix elements
display a strong scale dependence  in quantum chromodynamics (QCD) already at one-loop order
(see Ref.~\cite{Broadhurst:1994se} and references therein).
Throughout the paper, we quote estimates and bounds 
for the $\epsilon_i$ ($\tilde \epsilon_i$) at the renormalization 
scale  $\mu=2$~GeV in the $\overline{\rm MS}$ scheme, unless otherwise specified.
\end{itemize}

The Lagrangian of Eq.~(\ref{eq:leff10}) mediates all leading, low-energy charged-current weak
processes involving up and down quarks.  
In some charged-current  processes involving first-generation quarks
the theoretical and experimental precision has reached, 
or will soon reach, a level that allows
stringent bounds on new-physics effective couplings.
To set the stage for this discussion, we now provide an overview of
how the various BSM couplings of  Eq.~(\ref{eq:leff10}) can be probed experimentally---we 
explore these points in detail in the following sections. 
For context, we note that detailed expressions 
of the non-$(V-A)$ contributions to neutron and nuclear beta decay correlation coefficients 
 can be found in the papers by Jackson, Treiman,  and Wyld~\cite{ Jackson1957zz,Jackson1957206}, where 
one can re-express the Lee-Yang couplings~\cite{Lee:1956qn} they employ in 
terms of the  $\epsilon_\alpha$ and  $\tilde{\epsilon}_\beta$ 
using the expressions given in Eqs.~(\ref{eq:conversions}) below.

\begin{itemize}
\item
The combinations $(\epsilon_L \pm \epsilon_R)$ affect  the overall normalization of the
effective Fermi constant  in processes mediated by the vector and axial-vector current, respectively.
As discussed below, the hadronic matrix elements of the vector current are
known very precisely up to corrections due to QCD flavor symmetry breaking, that is, quark
mass differences, 
whereas the axial-vector matrix elements require non-perturbative calculations.
Therefore,  while the difference  $(\epsilon_L-\epsilon_R)$ remains relatively unconstrained,
the sum $(\epsilon_L+\epsilon_R)$ is strongly constrained by
quark-lepton universality tests, which are tantamount to CKM unitarity tests.
These tests involve a precise determination of 
$V_{ud}$  and $V_{us}$ from processes mediated by the vector current, 
such as  $0^+ \to 0^+$ nuclear decays and $K \to \pi \ell \nu$. 
An extensive analysis of the constraints on $(\epsilon_L+\epsilon_R)$
from universality tests and precision electroweak observables at  the
$Z$-pole was performed in Ref.~\cite{Cirigliano:2009wk}. 
In this context it was shown that constraints from low-energy are at the same level or
stronger---depending on the operator---than those from $Z$-pole observables and
$e^+ e^- \to q \bar{q}$ cross-section measurements at LEP.

\item
The right-handed coupling $\epsilon_R$ affects the relative normalization of
the axial and vector currents. 
In all beta decays $\epsilon_R$ can be
absorbed in a redefinition of the axial coupling, and,  
up to calculable radiative corrections~\cite{Czarnecki:2004cw,Shann:1971fz,Yokoo:1973mz,Garcia:1978bq,Gluck:1992qy, Fukugita:2004pq},
experiments  determine the combination $(1 - 2 \epsilon_R)g_A/g_V$, where $g_V$ and
$g_A$ are the vector and axial form factors at zero momentum transfer, to be 
precisely defined below.  Disentangling $\epsilon_R$ requires
precision measurements of $(1 - 2 \epsilon_R)g_A/g_V$ and precision
calculations of $g_A/g_V$ in lattice QCD, which, unfortunately, are not
yet at the required sub-percent level.

\item
The effective pseudoscalar coupling  $\epsilon_P$ 
contributes to the leptonic decays of the pion.  It is strongly
constrained by the helicity-suppressed ratio $R_\pi \equiv \Gamma(\pi \to e \nu
[\gamma])/\Gamma(\pi \to \mu \nu [\gamma])$.  Moreover, as discussed
in Refs.~\cite{Voloshin:1992sn,Herczeg:1994ur, Campbell:2003ir}, the
low-energy coupling $\epsilon_P$ receives contributions proportional
to $\epsilon_{S,T}$ through electroweak radiative corrections.

\item
Both the scalar  and tensor couplings  $\epsilon_S$ and  $\epsilon_T $ contribute at  linear
order  to the Fierz interference
term $b$ in the beta decays of neutrons and nuclei, as well as to the neutrino-asymmetry
correlation coefficient $B$ in polarized neutron and nuclear decays.
The empirical determination of the beta-asymmetry correlation
coefficient $A$ and 
the electron-neutrino correlation $a$ in neutron and nuclear beta decays, as 
well as  positron polarization measurements therein, entrain sensitivity to the
Fierz interference term as well. Thus 
bounds on $\epsilon_S$ and $\epsilon_T$ can also be obtained from these observables. 
Moreover,  the quadratic dependence on these couplings is  useful
in limiting their imaginary parts as well. 
Finally, the tensor coupling $\epsilon_T$ can also be constrained through 
Dalitz-plot studies of the radiative pion decay $\pi \to e \nu \gamma$.

\item
Neglecting neutrino masses, all the $\tilde{\epsilon}_\beta$ couplings contribute
to decay rates as per $\propto | \tilde{\epsilon}_\beta|^2$, 
so that it is more challenging to set limits on their appearance at low energies.

\item
All of the operators of Eq.~(\ref{eq:leff10}) can produce collider  signatures.
Before the advent of the LHC, 
collider bounds on the chirality-flipping scalar and tensor couplings
$\epsilon_{S,P,T}$ and  $\tilde{\epsilon}_{S,P,T}$ were very weak,
because their interference with the SM amplitude appears with 
factors of $m_f/E_f$, where $m_f$ is a light fermion mass with 
$f \in \{e,u,d\}$, which at collider energies strongly suppresses the whole effect. 
At the LHC, however, the contributions which appear as $|{\epsilon}_\beta|^2$ or 
$| \tilde{\epsilon}_\beta|^2$  can be boosted by 
a factor involving the energy in the numerator, noting that 
we replace $(v/\Lambda_{\rm BSM})^4  \to  (E/\Lambda_{\rm BSM})^4$, thus 
increasing the sensitivity to these couplings. 
We will discuss these bounds and show that 
with higher center-of-mass energy and integrated luminosity
they become  competitive with low-energy searches for $\epsilon_{S,T}$ 
or  stronger than low-energy bounds for $\tilde{\epsilon}_{R,S,T}$.
This analysis, of course, makes sense only  for 
$\Lambda_{\rm BSM} \gsim$~few TeV. 

\end{itemize}
The above considerations and more are summarized in Table~\ref{tab:probes}.

\begin{table}[h]
\begin{center}
\begin{minipage}[t]{16.5 cm}
\caption{
Summary of  the  most sensitive direct  low-energy probes of non-standard charged-current couplings.
Left column: combination of couplings. Right column: probe. 
The effective couplings are defined in Eq.~(\ref{eq:leff10}).  
The decay parameters $a,b,B,A$ are defined in Eq.~(\ref{eq:correlations}).
If  the new interactions originate at mass scales above the TeV, the LHC 
provides 
constraints on  all  
non-standard couplings through the process $pp \to e + \nu + X$. 
\label{tab:probes}}
\end{minipage}
\begin{tabular}{|c|c|}
\hline
 & \\
      Non-standard coupling              &         Probe     \\  
 & \\
\hline
\hline
 & \\
      $\epsilon_L + \epsilon_R$             &         CKM unitarity  \\  
  & \\    
\hline
 & \\
    $\epsilon_L -  \epsilon_R,  \  \ \epsilon_P,  \ \ \tilde{\epsilon}_P$          &        $R_\pi$ \\ 
  & \\    
\hline
 & \\
    $\epsilon_S$          &        $b, \  B \ \  [ a, \ A ]$   \\  
 & \\    
\hline            
 & \\
    $\epsilon_T$          &        $b, \ B \ \ [ a, \ A ], \ \ \pi \to e \nu \gamma$   \\  
 & \\    
\hline            
 & \\
    $\tilde{\epsilon}_{\alpha \neq P}$
      &       $a,b,B,A$   \\  
 & \\               
\hline
\end{tabular}
\end{center}
\end{table}

\subsection{\it  Hadronic  and nuclear matrix elements}

Hadronic and nuclear transition amplitudes always involve products of
short-distance couplings,  
evolved to the appropriate matching scale,  
and hadronic matrix elements. Thus 
in order to extract information on the former, 
we need to know the latter. 
Specifically, 
we need to match the quark-level effective theory of Eq.~(\ref{eq:leff10}) to
a low-energy effective theory written in terms of meson and baryon degrees of freedom. 
In QCD, this effective theory is 
Chiral Perturbation Theory (ChPT)~\cite{Weinberg:1978kz,Gasser:1983yg,Gasser:1984gg}. 
In the baryon sector, the low-energy structure of the theory is more complicated, 
and heavy baryon chiral perturbation theory is employed~\cite{Jenkins:1990jv}, 
where we refer the reader to Ref.~\cite{Bernard:1995dp} for a review. 
Different systematic approaches to remedy its limitations have been developed, 
improving the theory's convergence, 
notably the ``small scale expansion'' of 
Refs.~\cite{Hemmert:1996xg,Hemmert:1997tj,Hemmert:1997ye,Bernard:1998gv}, 
as well as Ref.~\cite{Becher:1999he}. 
As we have discussed, the precision with which we know 
the matrix elements of the SM operators limits our ability to constrain
new physics. 
If we wish to probe scales such that
$(v/\Lambda_{\rm BSM})^2 \sim 10^{-3}$, we need to know 
the SM matrix elements with commensurate
precision. 
This requires including all of the 
electromagnetic, isospin-breaking, and recoil-order effects in the calculation.
Since the operators appearing in Eq.~(\ref{eq:leff10}) 
have the factorized structure $J_{\rm quark} \times J_{\rm lepton}$,
we need not present the ChPT framework but rather 
can describe the purely hadronic effects in terms of 
meson and nucleon 
matrix elements of quark bilinears. 
Nevertheless,  the full ChPT machinery 
should ultimately be employed to compute long-distance radiative
corrections. In the case of neutron decay, this has been done in 
Ref.~\cite{Ando:2004rk},  finding  results consistent with 
non-ChPT based calculations~\cite{Czarnecki:2004cw,Shann:1971fz,Yokoo:1973mz,Garcia:1978bq,Gluck:1992qy, Fukugita:2004pq,
Ivanov:2012qe}. 
In this review, we will not  
further discuss long-distance radiative corrections to neutron decay  and refer the reader to 
Refs.~\cite{Ando:2004rk} and \cite{Czarnecki:2004cw} for recent detailed accounts.

\subsubsection{\it Meson matrix elements}

Leptonic ($M \to l  \nu$) and semi-leptonic ($M_1 \to M_2 l  \nu$) decays of 
pseudoscalar mesons provide strong constraints on the CC BSM couplings.
The relevant  one-meson matrix elements are parameterized in terms of the pion 
and kaon decay constants $F_{\pi,K}$ as follows (in our normalization $F_\pi \simeq 92$~MeV):
\bea
\langle 0 | \bar{u} \gamma_\mu \gamma_5  d   |  \pi^- (p)  \rangle&=&  - i \sqrt{2}  \, F_\pi  \  p_\mu
\\
\langle 0 | \bar{u} \gamma_\mu \gamma_5  s   |  K^- (p)  \rangle &=&  - i \sqrt{2}  \,   F_K \    p_\mu
\\
\langle 0 | \bar{u}  \gamma_5  d   | \pi^- (p)  \rangle &=& i  \  \frac{m_\pi^2}{m_u + m_d} \,  \sqrt{2}  \,  F_\pi
\\
\langle 0 | \bar{u}  \gamma_5  s   | K^- (p)  \rangle &=& i  \  \frac{m_K^2}{m_u + m_s} \, \sqrt{2}  \,  F_K \,.
\eea
The pseudoscalar matrix elements follow from the axial-vector  ones by using the operator
relation 
$\partial_\mu  \, \bar{q}_i  \gamma^\mu \gamma_5  q_j  \ = \  i (m_i + m_j) \   \bar{q}_j \gamma_5 q_i$.
Matrix elements of the other bilinears vanish by parity.

The two-meson matrix elements of vector and scalar densities 
can be parameterized in terms of two form factors.
Specializing to the $K^0 \to \pi^-$ transitions, we have, noting the momentum transfer 
$q = p-k$:
\bea
\langle \pi^-(k)  | \bar{s} \gamma_\mu  u    |  K^0  (p)  \rangle \  &=&  \  (p + k)_\mu  \, {  f_+ (q^2) }  \ + \
 (p - k)_\mu  \ { f_- (q^2) }
\\
\langle \pi^-(k)  | \bar{s}   u    |  K^0  (p)  \rangle \  &=&   - \frac{m_K^2 - m_\pi^2}{m_s - m_u} \ {f_0 (q^2) } \,,
\eea
where
$f_0 (q^2) = f_+ (q^2)  +  ({q^2}/({m_K^2 - m_\pi^2})) f_- (q^2)$
and  the operator relation
$\partial_\mu \, \bar{q}_i  \gamma^\mu q_j  \ = \  i (m_i - m_j) \   \bar{q}_j q_i$
has been used to relate the vector and scalar matrix elements.
In the SU(3)$_f$ limit the light quark masses obey $m_u=m_d=m_s$, 
so that $f_+(0)=1$---the Ademollo-Gatto theorem~\cite{Ademollo:1964sr,Behrends:1960nf}
ensures that the corrections to the flavor symmetry limit start at second order:
 $f_+(0)=1 + {\cal O}( (m_s - m_d)^2)$.\footnote{Note, however, that in the case of 
charged $K_{\ell 3}$ decay the existence of $\pi^0-\eta\,,\eta'$ 
mixing implies that corrections to $f_+(0)=1$ occur 
at first order in $(m_d-m_u)/(m_s-\hat{m})$, with $\hat{m}=(m_d + m_u)/2$.}
Finally, for completeness, 
we report the tensor matrix element, which involves a new dynamical form factor $B_T (q^2)$~\cite{Becirevic:2000zi}:
\be
\langle \pi^-(k)  | \bar{s} \sigma_{\mu \nu}   u    |  K^0  (p)  \rangle \  =
i \ \frac{p_\mu k_\nu - p_\nu k_\mu}{m_K}   \, { B_T (q^2) } ~.
\ee

The decay constants and form-factors can be calculated in 
lattice QCD (LQCD) and we will review the relevant results as needed.

\subsubsection{\it Nucleon matrix elements}

At the one-nucleon level, we require the matrix elements
between the neutron and proton
of all possible quark bilinears of dimension three. These
can be parameterized in terms of Lorentz-invariant form factors as
follows~\cite{Weinberg:1958ut}:
\begin{subequations}
\label{eq:nucleonmatching}
\bea
\br{p (p_p) } \bar{u} \gamma_\mu d \ket{n (p_n)} &=&
\bar{u}_p (p_p)  \left[
g_V(q^2)  \,  \gamma_\mu
- i \,  \frac{\tilde{g}_{T(V)} (q^2)}{2 M_N}   \, \sigma_{\mu \nu}   q^\nu
+ \frac{\tilde{g}_{S} (q^2)}{2 M_N}   \,  q_\mu
\right]
\hspace{-0.1cm}
 u_n (p_n) \\&&
\nonumber \\
\br{p (p_p) } \bar{u} \gamma_\mu \gamma_5  d \ket{n (p_n)} &=&
\bar{u}_p (p_p)  \left[
g_A(q^2)    \gamma_\mu
\hspace{-0.1cm}
- i \,  \frac{\tilde{g}_{T(A)} (q^2)}{2 M_N}   \sigma_{\mu \nu}   q^\nu
+
\hspace{-0.1cm}
\frac{\tilde{g}_{P} (q^2)}{2 M_N}   q_\mu
\right]  \hspace{-0.15cm}  \gamma_5  u_n (p_n)  
\\ \label{eq:inducedgP}
\nonumber &&
\\
\br{p (p_p) } \bar{u} \,   d \ket{n (p_n)} &=&
g_S(q^2)  \ \bar{u}_p (p_p)  \, u_n (p_n)
\\
\br{p (p_p) } \bar{u} \,  \gamma_5 \,  d \ket{n (p_n)} &=&
g_P(q^2)  \ \bar{u}_p (p_p)  \, \gamma_5 \, u_n (p_n)
\label{eq:defgP}
\\
\br{p (p_p) } \bar{u} \, \sigma_{\mu \nu}  \,  d \ket{n (p_n)} &=&
 \bar{u}_p (p_p)
 \left[
g_T(q^2) \, \sigma_{\mu \nu}   +  g_{T}^{(1)} (q^2)  \left(q_\mu \gamma_\nu - q_\nu \gamma_\mu \right)
\nonumber  \right. \\
&+&
\left.  g_{T}^{(2)} (q^2)  \left( q_\mu P_\nu - q_\nu P_\mu  \right)
+
g_{T}^{(3)} (q^2)
 \left(
\gamma_\mu  \slashed{q}  \gamma_\nu -
\gamma_\nu  \slashed{q} \gamma_\mu
 \right)
\right]
u_n (p_n) \,,
\eea
\end{subequations}
where $u_{p,n}$ are the proton and neutron spinors,   $P = p_n + p_p$,
$q = p_n - p_p$ is the momentum transfer,  and $M_N= (M_n + M_p)/2$ denotes a common nucleon
mass.\footnote{In the case of vector and axial bilinears, the Gordon decomposition
can be used to trade the induced tensor term proportional to
$\sigma_{\mu \nu} q^\nu$ for an independent scalar term proportional to $P_\mu$.
Here we choose to follow the parameterization of Ref.~\cite{Weinberg:1958ut}.}
Note that the above spinor contractions are ${\cal O}(1)$, 
except for $\bar{u}_p \gamma_5 u_n $, which is ${\cal O}(q/M_N)$.

In order to make contact with the standard
references on neutron and nuclear beta-decay
phenomenology~\cite{Lee:1956qn,Jackson1957zz,Jackson1957206,Severijns2006dr}, we
note that upon 
neglecting recoil order terms
Eq.~(\ref{eq:nucleonmatching}) can be viewed as the
matching conditions from our quark-level effective
theory Eq.~(\ref{eq:leff10}) to the 
nucleon-level effective theory
originally written down by Lee and Yang~\cite{Lee:1956qn}.
The Lee-Yang  effective couplings $C_i$, $C_i'$ ($i \in \{V,A,S,T\}$) can be
expressed in terms of our parameters as follows~\cite{Cirigliano:2012ab}~\footnote{
Various metrics and conventions appear in the literature. Lee and Yang~\cite{Lee:1956qn}
employ the ``ict'' metric, which in this case 
maps to the metric 
we employ if we let $\gamma_5 \to - \gamma_5$ in their effective theory, 
noting $\gamma_5 = i \gamma^0 \gamma^1 \gamma^2 \gamma^3$. Refs.~\cite{Holstein:1974zf}
and \cite{Severijns:2011zz} 
employ the metric we do but flip the sign of the $\gamma_5$ terms. 
}
\begin{subequations}
\bea
C_{i} &=& \frac{G_F^{(0)}}{\sqrt{2}} \,  V_{ud} \,  \bar{C}_{i}    \\
\bar{C}_V &=& g_V  \left(1 +  \delta_{\beta} + \epsilon_L + \epsilon_R
+ \tilde{\epsilon}_L + \tilde{\epsilon}_R
\right)  \\
\bar{C}_V' &=& g_V  \left(1 + \delta_{\beta} + \epsilon_L + \epsilon_R
- \tilde{\epsilon}_L - \tilde{\epsilon}_R
\right)  \\
\bar{C}_A &=& - g_A  \left(1 + \delta_{\beta} + \epsilon_L - \epsilon_R
- \tilde{\epsilon}_L   +  \tilde{\epsilon}_R
\right)  \\
\bar{C}_A' &=& - g_A  \left(1 + \delta_{\beta} + \epsilon_L - \epsilon_R
+ \tilde{\epsilon}_L   -  \tilde{\epsilon}_R
\right)  \\
\bar{C}_S &=&   g_S  \, \left(  \epsilon_S  + \tilde{\epsilon}_S \right) \\
\bar{C}_S'  &=&   g_S  \, \left(  \epsilon_S  -  \tilde{\epsilon}_S \right) \\
\bar{C}_P &=&   g_P  \, \left(  \epsilon_P  - \tilde{\epsilon}_P \right) \\
\bar{C}_P'  &=&   g_P  \, \left(  \epsilon_P  +  \tilde{\epsilon}_P  \right) \\
\bar{C}_T &=&    4 \, g_T \, \left( \epsilon_T   + \tilde{\epsilon}_T \right)  \\
\bar{C}_T '  &=&    4 \, g_T \, \left( \epsilon_T   -  \tilde{\epsilon}_T \right) ~.
\eea
\label{eq:conversions}
\end{subequations}
Using these relations and the results of Ref.~\cite{Jackson1957zz}
one can easily work out the dependence of beta decay observables on the
short-distance parameters $\epsilon_i$ and $\tilde{\epsilon}_i$.

Our goal is to identify TeV-scale induced 
new physics 
contaminations 
of typical size
$\epsilon_{\alpha} \sim (v/\Lambda_{\rm BSM})^2  \sim {\cal O}(10^{-3})$
to the decay amplitude, so that they are comparable in size to the
recoil corrections of ${\cal O}(q/M_N) \sim 10^{-3}$ and 
the radiative corrections of ${\cal O}(\alpha/\pi)$.
Thus, it is useful to organize the discussion in terms of a 
simultaneous expansion in new physics contributions, recoil, 
and radiative corrections keeping terms through first order only.
Higher-order terms may prove negligible in light of 
anticipated experimental sensitivities, but 
we indicate the role of certain, more significant ones. 
Employing this simultaneous expansion in $\epsilon_{\alpha}$, $q/M_N$, and $\alpha/\pi$,
we now discuss  the  contributions from the quark-bilinear operators:

\begin{itemize}
\item  {\bf Vector current}:
The form factor $g_V(0)$ contributes at ${\cal O}(1)$ to the amplitude, whereas 
$\tilde{g}_{T(V)} (0)$ and $\tilde{g}_{S} (0)$ 
contribute at ${\cal O}(q/M_N)$. Also, in the SU(2)$_f$, or isospin, 
limit, the weak magnetism form factor $\tilde{g}_{T(V)} (0)$ is related
to the difference of the empirical proton and neutron magnetic moments, which are 
well-known, and the induced-scalar form factor $\tilde{g}_S (q^2)$, reflective of
the presence of a second-class current, vanishes~\cite{Weinberg:1958ut}. 
Corrections to the isospin limit are of ${\cal O}((M_n - M_p)/M_N) \sim q/M_N$. 
Since $\tilde{g}_S$  multiplies one power of $q_\mu/M_N$, 
its contribution to the decay amplitude is effectively of 
second order in the recoil expansion.

\item  {\bf Axial current}: 
The form factor $g_A (0)$ contributes at ${\cal O}(1)$, whereas  
$\tilde{g}_{T(A)} (0)$ and $\tilde{g}_{P} (0)$ 
contribute at ${\cal O}(q/M_N)$. 
The induced-tensor form factor $\tilde{g}_{T(A)}(q^2)$ vanishes in the 
isospin limit~\cite{Weinberg:1958ut}, so that its contribution to the decay amplitude
is of second order in $q/M_N$. Similarly, the
contribution associated with the induced-pseudoscalar form factor
$\tilde{g}_P$ is quadratic in our counting, because the pseudoscalar
bilinear is itself of order $q/M_N$, and is accompanied by an explicit
$q/M_N$ suppression \footnote{This term, however, is 
enhanced. Using the partially conserved axial current one can show that
the form factor $\tilde{g}_P$ is of order $M_N/m_q \sim 100$, making a 
${\cal O}(10^{-4})$ contribution to the amplitude. The effect of
$\tilde{g}_P$ on  beta-decay rates has been worked out in
Ref.~\cite{Holstein:1974zf}, and it should be included when the
experiments reach that level of precision.}; 
it can be studied in muon capture, note Ref.~\cite{Gorringe:2002xx,Bernard:2001rs}
for reviews.

\item  {\bf Pseudoscalar bilinear}:
The pseudoscalar bilinear $\bar{u}_p \gamma_5 u_n$ is itself of order $q/M_N$.
Since it necessarily multiplies a BSM effective coupling $\epsilon_P$ because 
there is no pseudoscalar coupling in the SM, this term is also of
second order in our expansion.

\item {\bf Scalar and tensor bilinears}:
These  bilinears enter into the analysis multiplied by 
new-physics effective couplings $\epsilon_{S,T}$.
Computing 
the corresponding matrix elements to zeroth order in the recoil expansion suffices 
to identify $g_{S}(0)$ and $g_T(0)$. Note that 
$g_T^{(1,2,3)}(q^2)$ appear only in ${\cal O}(q/M_N)$, 
and $g_T^{(3)}(q^2)$ vanishes in the isospin limit~\cite{Weinberg:1958ut}.

\end{itemize}

In summary,  to the order we are working, the amplitudes depend only
on $g_i \equiv g_i(0)$ with $i \in \{ V,A,S,T \}$ and $\tilde{g}_{T(V)} (0)$.
Up to second-order corrections in isospin breaking, 
one has $g_V = 1$~\cite{Ademollo:1964sr,Donoghue:1990ti,Kaiser:2001yc}.
We define the ratio of the
axial to vector form factors as $\lambda \equiv g_A/g_V$, where $\lambda>0$ under our 
conventions.  As we have noted, 
the neutron-decay amplitude in the presence of 
non-standard right-handed interactions is actually a
function of $\tilde{\lambda} \equiv \lambda (1 - 2 \epsilon_R)$. The parameter
$\tilde{\lambda}$ is extracted very precisely from beta-asymmetry measurements in
polarized neutron decay,  leading to  
$\tilde{\lambda} =  1.2701(25)$~\cite{Beringer:2012zz}; this number 
is essentially $g_A$. There are no direct experimental handles on $g_{S,P,T}$. 

A first principles  calculation of $g_{A,S,P,T}$, however, is possible with LQCD. 
The status of LQCD calculations of these charges is critically reviewed 
in~\cite{Bhattacharya:2011qm}, and in which 
the first estimate of $g_S$  from LQCD is provided.
Different calculations give results in the range $1.12 < g_A < 1.26$; 
the errors are much larger than the experimental uncertainty.
In constrast, the new estimates for the  scalar  and tensor charge
in the $\overline{MS}$ scheme and at $\mu=2$~GeV
are  $g_S = 0.8 \pm 0.4$ and $g_T = 1.05 \pm 0.35$.
Besides statistical uncertainties, which are particularly large for $g_S$, 
the dominant LQCD systematic effects 
in $g_{S,T}$ arise from extrapolation in the quark mass to the physical point
and from the renormalization constants, noting 
perturbative calculations were used to arrive at the reported results, and 
the non-perturbative calculation is in progress.
More recently Ref.~\cite{Green:2012ej} provided improved results for the scalar
and tensor charges,  $g_S = 1.08 \pm 0.28$ and $g_T = 1.038 \pm 0.011$,
with the uncertainty associated with statistics and chiral extrapolation.
These results, however, do not include an estimate of the
systematic error associated with finite volume and finite lattice spacing extrapolations.
Therefore, in what follows we use the results of Ref.~\cite{Bhattacharya:2011qm}
as the baseline lattice results.

\subsubsection{\it  Nuclear matrix elements}

In moving from neutron decay to a general nuclear beta decay, 
a number of important differences appear. 
For one thing, numerous spin sequences are possible. Also 
the daughter state may well itself be unstable 
under electromagnetic or strong interactions. Finally, 
the $Q$-value, defined as the maximum electron (positron) kinetic energy, 
is generally much larger than the 0.8 MeV 
found in neutron beta decay and in some cases can be as large as 10-15 MeV. 
The electron/positron energy dependence of the decay observables are then much
more appreciable.

The spins are easily dealt with by the use of Clebsch-Gordan 
coefficients  $C^{M_1,M_2;M}_{J_1,J_2;J}$
together with the Wigner-Eckart theorem to define 
reduced matrix elements.  Thus the general form for a vector and axial-vector 
nuclear matrix element between parent and daughter states having
 spins $J,M$ and $J',M'$ and masses $M_1$ and $M_2$, 
respectively, is~\cite{Holstein:1974zf}\footnote{Here we 
discuss only allowed decays, for which $\Delta J=0,\pm 1$ with no change in nuclear parity.}
\begin{eqnarray}
\ell^\mu<\beta|V_\mu|\alpha>&=&\left(a(q^2){P\cdot\ell\over
2M_A}+e(q^2){q\cdot\ell\over
2M_A}\right)\delta_{JJ'}\delta_{MM'}
+ i{\tilde{b}(q^2)\over
2M_A}C_{J'1;J}^{M'k;M}(\vec{q}\times\vec{\ell})_k\nonumber\\
&+&C_{J'2;J}^{M'k;M}\left[{f(q^2)\over
2M_A}C_{11;2}^{nn';k}\ell_nq_{n'}
+  {g(q^2)\over (2M_A)^3}P\cdot\ell\sqrt{4\pi\over
5}Y_2^k(\hat{q})\vec{q}^{\,2}+\ldots\right]
\nonumber\\
\ell^\mu<\beta|A_\mu|\alpha>&=&C_{J'1;J}^{M'k;M}\epsilon_{ijk}\epsilon_{ij\lambda\eta}
{1\over 4M_A}\left[c(q^2)\ell^\lambda P^\eta-d(q^2)\ell^\lambda
q^\eta
+ {1\over (2M_A)^2}h(q^2)q^\lambda P^\eta
q\cdot\ell\right]\nonumber\\
&+&C_{J'2;J}^{M'k;M}C_{12;2}^{nn';k}\ell_n\sqrt{4\pi\over
5}Y_2^{n'}(\hat{q}){\vec{q}^{\,2}\over (2M_A)^2}j_2(q^2)\nonumber\\
&+&C_{J'3;J}^{M'k;M}C_{12;3}^{nn';k}\ell_n\sqrt{4\pi\over
5}Y_2^{n'}(\hat{q}){\vec{q}^{\,2}\over (2M_A)^2}j_3(q^2)+\ldots \,,
\end{eqnarray}
where $\ell^\mu$ denotes the leptonic current and $M_A=(M_1 + M_2)/2$. 
Here each term corresponds to one in the analogous neutron transition
via
\begin{equation}
\begin{array}{cc}
a\rightarrow g_V,&\quad c\rightarrow g_A \sqrt{3}\\
\tilde{b}\rightarrow \tilde{g}_{T(V)} \sqrt{3},&\quad d\rightarrow \tilde{g}_{T(A)} \sqrt{3}\\
e\rightarrow \tilde{g}_S,&\quad h\rightarrow \tilde{g}_P \sqrt{3} \,.
\end{array}
\end{equation}
In addition, there exist terms $f,g,j_2,j_3$ which have no $J={1\over
2}\rightarrow J'={1\over 2}$ analog since they involve $\Delta J=2,3$.

For each form factor there exist known one-body operator, or impulse
approximation, predictions.  Defining the nuclear mass difference $\Delta=M_1-M_2$,
we have
\begin{eqnarray}
a(q^2)&\simeq&(1+{\Delta\over 2M_A})^{-1}g_V(q^2)
\times [{\cal M}_F+{1\over 6}(q^2-\Delta^2){\cal
M}_{r^2}+{\Delta\over 3}{\cal M}_{r\cdot p}]\nonumber\\
\tilde{b}(q^2)&\simeq&A[\tilde{g}_{T(V)}(q^2){\cal M}_{GT}+g_V(q^2){\cal M}_{L}]\nonumber\\
c(q^2)&\simeq&(1+{\Delta\over 2M_A})^{-1}g_A(q^2)[{\cal M}_{GT}+{1\over
6}(q^2-\Delta^2){\cal M}_{\sigma r^2}
+ {1\over 6\sqrt{10}}{\cal M}_{1{\cal Y}}(2\Delta^2+q^2)\nonumber\\
&+&A{\Delta\over 2M_A}{\cal M}_{\sigma L}+{\Delta\over 2}{\cal M}_{\sigma rp}]\nonumber\\
d(q^2)&\simeq&(1+{\Delta\over 2M_A})^{-1}g_A(q^2)[-{\cal M}_{GT}-{1\over
6}(q^2-\Delta^2){\cal M}_{\sigma r^2}\nonumber\\
&+&{1\over \sqrt{10}}{\cal M}_{1{\cal Y}}(M_A\Delta +{1\over 6}(\Delta^2-q^2))\nonumber\\
&+&A{\cal M}_{\sigma L}+M_A{\cal M}_{\sigma rp}]\pm A\tilde{g}_{T(A)}(q^2){\cal
M}_{GT}\nonumber\\
e(q^2)&\simeq&(1+{\Delta\over 2M_A})^{-1}g_V(q^2)[{\cal M}_F+{1\over
6}(q^2-\Delta^2){\cal M}_{r^2}
- {2M_A\over 3}{\cal M}_{r\cdot p}]\pm A\tilde{g}_S(q^2){\cal M}_F\nonumber\\
f(q^2)&\simeq&g_V(q^2)2M_A{\cal M}_{\{r,p\}}\nonumber\\
g(q^2)&\simeq&-g_V(q^2){4M_A^2\over 3}{\cal M}_Q\nonumber\\
h(q^2)&\simeq&-(1+{\Delta\over 2M_A})^{-1}
\times [g_A(q^2){2M_A^2\over \sqrt{10}}{\cal M}_{1{\cal
Y}}+\tilde{g}_P(q^2)A^2{\cal M}_{GT}]\nonumber\\
j_i(q^2)&\simeq&-{2M_A^2\over 3}g_A(q^2){\cal M}_{i{\cal Y}},\quad i=2,3 \,,
\end{eqnarray}
where the upper (lower) sign refers to electron (positron) emission and 
 the ${\cal M}$'s represent reduced, nonrelativistic matrix elements, namely, 
\begin{eqnarray}
{\cal M}_F&=&<\beta||\sum_i\tau_i^\pm||\alpha>\nonumber\\
{\cal M}_{GT}&=&<\beta||\sum_i\tau_i^\pm\vec{\sigma}_i||\alpha>\nonumber\\
{\cal M}_{r^2}&=&<\beta||\sum_i\tau_i^\pm r_i^2||\alpha>\nonumber\\
{\cal M}_{\sigma
r^2}&=&<\beta||\sum_i\tau_i^\pm r_i^2\vec{\sigma}_i||\alpha>\nonumber\\
{\cal M}_{r\cdot p}&=&{i\over
2M_N}<\beta||\sum_i\tau_i^\pm(\vec{r}_i\cdot
\vec{p}_i+\vec{p}_i\cdot\vec{r}_i)||\alpha>\nonumber\\
{\cal M}_{\sigma L}&=&<\beta||\sum_i\tau_i^\pm\vec{\sigma}_i\times
(\vec{r}_i\times\vec{p}_i)||\alpha>\nonumber\\
{\cal M}_{K{\cal Y}}&=&\sqrt{16\pi\over
5}<\beta||\sum_i\tau_i^\pm r_i^2C_{12;K}^{nn';k}
\times \sigma_{in}Y_2^{n'}(\hat{r}_i)||\alpha>,\quad K=1,2,3\nonumber\\
{\cal M}_{\sigma rp}&=&{i\over 2M_N}<\beta||\sum_i\tau_i^\pm
[\{\vec{\sigma}_i\cdot\vec{r}_i,\vec{p}_i\}
+ \{\vec{\sigma}_i\cdot\vec{p}_i,\vec{r_i}\}]||\alpha> \,,
\end{eqnarray}
as well as 
\begin{eqnarray}
{\cal
M}_Q&=&<\beta||\sum_i\tau_i^\pm r_i^2Y_2^k(\hat{r}_i)||\alpha>\nonumber\\
{\cal M}_{\{r,p\}}&=&{i\over 2M_N}<\beta||\sum_i\tau_i^\pm
C_{11;2}^{nn';k}
\times (r_{in}p_{in'}+p_{in}r_{in'})||\alpha>\nonumber\\
{\cal M}_L&=&<\beta||\sum_i\tau_i^\pm(\vec{r}_i\times\vec{p}_i)||\alpha> \,.
\end{eqnarray}
One expects meson exchange, or two-body, corrections to
these predictions at the order of 5 to 10\% or so~\cite{Chemtob:1971pu}.

If we neglect recoil effects, which is generally a good approximation because
they enter the nuclear matrix elements at  ${\cal O}(\Delta/M_N)$, 
there exist only the leading Fermi and Gamow-Teller form factors $a=g_V{\cal M}_F$ 
and $c=g_A{\cal M}_{GT}$. Most experiments 
are analyzed in terms of only these quantities. 
In the SM the vector weak charge 
is also the isospin raising operator, so that the Fermi form factor 
$a(0)$ must vanish unless the parent and daughter states 
are isotopic analogs. Moreover, 
if the parent-daughter 
analog states have spin-parity $0^+$, then the Gamow-Teller form factor 
vanishes---$c(q^2)=0$---and the Fermi matrix element becomes a simple 
numerical factor, namely, for unit isotopic spin $a(0)=\sqrt{2}$, which is an exact 
prediction up to isospin breaking effects. 
The calculation of the latter effects has been an area of active 
investigation~\cite{Ormand:1995df,Hardy2004id, 
Auerbach:2008ut,Liang:2009pf,Satula:2010uz,
Miller:2008my,Miller:2009cg,Towner:2010bx}.
In simple terms, one needs to take into account  that 
the ``last" proton in the parent positron emitter is less strongly bound than 
the ``last" neutron in the daughter state.
An extensive analysis of data by Hardy and Towner, taking into account isospin breaking,  has  
yielded the very precise value $V_{ud}=0.97425(22)$~\cite{Towner:2010zz}. 
%

No symmetry principle determines 
the size of Gamow-Teller matrix elements, and these must 
either be determined empirically from lifetime or correlation coefficient 
measurements or calculated from nuclear wavefunctions, 
though the latter can only be done precisely if meson exchange effects are included. 
In the case of BSM matrix elements, 
wavefunction calculations of the requisite matrix elements are required.

\section{Cabibbo and lepton universality
 \label{sect:universality}}

Beta decays provide stringent constraints on  non-standard couplings through two
classes of  observables.
(i) Total decay rates,   after inclusion of radiative corrections,
provide information on the  strength of weak interactions, thus enabling
precision tests of 
Cabibbo  and lepton universality.
(ii)  Differential decay distributions, including spectra and correlations, 
are sensitive to the Lorentz structure of the underlying weak interaction,
thus enabling searches for small non-$(V-A)$  components. 
In this section we focus on universality tests.

\subsection{\it Cabibbo Universality}
\label{sect:CKM}

In the SM the effective 
Fermi constant $G_\beta$ controlling 
semi-leptonic transitions $u_i \leftrightarrow d_j$ 
for $i,j \in 1,2,3$ 
is related to $G_\mu$  by  
\be
G_\beta = G_\mu \, V_{ij}  \ \Big(1 + \delta_\beta - \delta_\mu \Big) \,,
\ee
where $G_\beta \equiv V_{ij} G_F^{(0)}(1 + \delta_\beta)$, 
the unitary CKM matrix $V_{ij}$ parameterizes quark mixing, 
and $\delta_\beta$ and $\delta_\mu$ encode electroweak 
radiative corrections (see Eqs.~(\ref{eq:leff10}) and (\ref{eq:leffmu})). 
We can thus test Cabibbo (or quark-lepton) universality by testing whether 
$|V_{ud}|^2 + |V_{us}|^2 + |V_{ub}|^2 =1$.
Beta decay rates permit access to the CKM matrix elements
$V_{ud}$ and $V_{us}$. Since both  the SM prediction and the experimental
measurements have reached the sub-percent level, these observables
provide strong constraints on new physics, through the parameter
$\Delta_{\rm CKM}$ defined as
\be
\label{eq:dckm}
\Delta_{\rm CKM} \equiv  |\overline{V}_{ud}|^2+|\overline{V}_{us}|^2+|\overline{V}_{ub}|^2 \ - \ 1 ~. 
\ee
Here $\overline{V}_{ij}$ 
are the CKM elements  determined phenomenologically from semileptonic decays assuming only SM dynamics. 
The ratio $\overline{V}_{ij}/V_{ij}$ is parameterized in terms of the BSM couplings, as exemplified by Eqs.~(\ref{eq:Vud1}) and (\ref{eq:Vudn}).
In the unitarity sum, 
$|\overline{V}_{ub}| = 3.51 ({\buildrel +15 \over {_{-14}}}) \times 10^{-3}$~\cite{Beringer:2012zz}  
plays no role; $\overline{V}_{ud}$  and $\overline{V}_{us}$ are both  important and 
can be determined with high precision in a number of channels.
The degree of  needed theoretical input varies, depending on the component of the weak
current which contributes to the hadronic  matrix element.  Roughly speaking,
one can group the channels leading to $\overline{V}_{ud,us}$ into three classes:
\begin{itemize}
\item  Semileptonic decays in which only the vector component of the weak current contributes:
These are theoretically favorable in the SM because the matrix elements of
the vector current at zero momentum transfer are known in the SU(2)$_f$ (SU(3)$_f$) limit of
equal light quark masses.
Moreover, corrections to the symmetry limit are quadratic in
$m_{s,d} - m_u$~\cite{Behrends:1960nf,Ademollo:1964sr}.
Super-allowed nuclear beta decays ($0^+ \to 0^+$), 
pion beta decay  ($\pi^+ \to \pi^0 e^+  \nu_e$), and
$K \to \pi \ell \nu$ decays belong to this class.
The determination of $\overline{V}_{ud,us}$ from these modes requires theoretical input
on radiative corrections~\cite{Marciano:1985pd,Czarnecki:2004cw,Marciano:2005ec,Cirigliano:2001mk,Cirigliano:2004pv,Cirigliano:2008wn}
and hadronic matrix elements via analytic methods~\cite{Hardy:2008gy,Leutwyler:1984je,Bijnens:2003uy,Jamin:2004re,Cirigliano:2005xn,Hill:2006bq,Bernard:2009zm},
or lattice QCD methods~\cite{Becirevic:2004ya,Dawson:2006qc,Boyle:2007qe,Lubicz:2009ht,Boyle:2010bh}.

\item Semileptonic transitions in which both the vector and axial component of the weak current contribute:
Neutron decay ($n \to p e \bar{\nu}$) and hyperon decays ($\Lambda \to p e \bar{\nu}$, ....),
as well as nuclear mirror transitions,  belong to this class.
In this case the matrix elements of 
the axial current must be determined 
experimentally~\cite{Cabibbo:2003cu,NaviliatCuncic:2008xt}.
Inclusive $\tau$ lepton decays 
$\tau \to h \nu_\tau$ also belong to this class as both $V$ and $A$ currents 
contribute, though 
the relevant matrix elements can be calculated theoretically via the Operator Product
Expansion~\cite{Braaten:1991qm,Gamiz:2002nu}.

\item Leptonic transitions in which only the axial component of the weak current contributes:
In this class one finds meson decays such as $\pi (K)  \to \mu \nu$ but also
exclusive $\tau$ decays such as $\tau \to \nu_\tau \pi (K)$.
Experimentally one can determine the products $\overline{V}_{ud} F_\pi$ and $\overline{V}_{us}  F_K$.
With the advent of precision calculations of $F_K/F_\pi$ in 
lattice QCD~\cite{Aubin:2004fs,Beane:2006kx,Follana:2007uv,Blossier:2009bx,Bazavov:2009bb}, 
this  class of decays provides a useful constraint on the ratio
$\overline{V}_{us}/\overline{V}_{ud}$~\cite{Marciano:2004uf,Cirigliano:2011tm}.

\end{itemize}

Currently, the determination of $\overline{V}_{ud}$ is 
dominated by $0^+ \to 0^+$ super-allowed nuclear
beta decays~\cite{Hardy:2008gy,Towner:2010zz}, leading to 
$\overline{V}_{ud} = 0.97425(22)$, 
while  the best determination 
of $\overline{V}_{us}$ arises from $K_{\ell 2},K_{\ell 3}$ decays, leading to the best fit
$\overline{V}_{us} = 0.2256(9)$~\cite{Antonelli:2010yf,Cirigliano:2011tm}.
These determinations lead to $\Delta_{\rm CKM} = (1 \pm 6) \times 10^{-4}$, 
in remarkable agreement with the CKM unitarity of the SM.
This is illustrated in Fig.~\ref{fig:CKM}.
Next, we discuss the implications on BSM physics in a model-independent framework.

\begin{figure}[t!]
\centering
\includegraphics[width=0.7\textwidth]{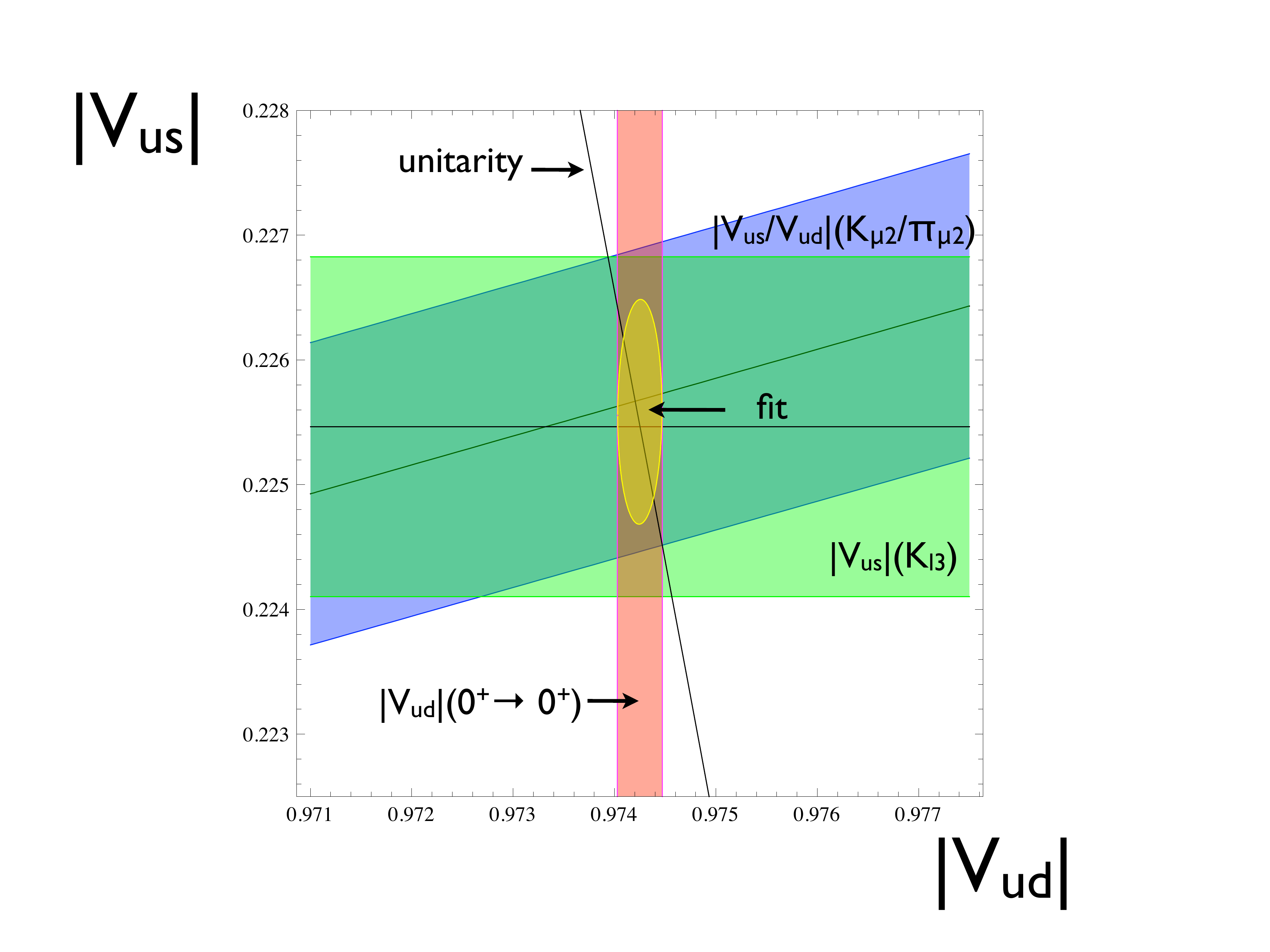}
\begin{minipage}[t]{16.5 cm}
\caption{
Graphical representation of the current status
of $|V_{ud}|$, $|V_{us}|$, and the corresponding CKM unitarity test.
The horizontal band represents the constraint from $K_{\ell 3}$ decays,
the thin vertical band the constraint from $0^+ \to 0^+$ nuclear decays,
the oblique band the constraint from $K_{\mu 2}/\pi_{\mu 2}$, whereas 
the ellipse is the $1 \sigma$  fit region. Figure reproduced with permission  from Ref.~\cite{Cirigliano:2011tm}.
}
\label{fig:CKM}
\end{minipage}
\end{figure}

\subsubsection{\it Model independent constraints}
\label{sec:cabibbo1}

Each element  $\overline{V}_{ij}$ receives a universal, or channel independent, correction
due to possible new physics corrections in muon decay,  parameterized by 
$\epsilon_\mu$ in Eq.~(\ref{eq:leffmu}). 
Additionally,  $\overline{V}_{ij}$ receives channel-dependent BSM contributions, which 
are linear combinations of the $\epsilon$'s defined in Eq.~(\ref{eq:leff10}).
Given the hierarchy $|V_{ud}|^2 \gg |V_{us}|^2$,  we discuss in detail
only  the BSM contributions to  $\overline{V}_{ud}$, 
beginning with the $0^+ \to 0^+$ nuclear transitions.

From each $0^+ \to 0^+$ transition, working in the impulse approximation, 
one extracts the quantity
\be
|\overline{V}_{ud}|^2_{0^+ \to 0^+}
= |V_{ud}|^2  \  \Big[  1  +  2 \, {\rm Re} ( \epsilon_L  +  \epsilon_R  - \epsilon_\mu)
+  c_{0^+}^{S} (Z)   \ g_S \, {\rm Re}\, \epsilon_S \ \Big] ~,
\label{eq:Vud1}
\ee
where the first correction reflects the BSM shift in the vector operator
minus  the shift in the Fermi constant extracted in muon decay.
The second correction, proportional to $\rm{Re}\,\epsilon_S$, arises from a non-vanishing
Fierz interference term; this distorts 
the electron spectrum and therefore the phase space integals.
The correction depends on the individual nuclear transitions,
through~\cite{Bhattacharya:2011qm,Czarnecki:2004cw}
\be
c_{0^+}^{S} (Z) = - 2 \sqrt{1 - \alpha^2 Z^2}  \,  \frac{I_1 (Q_{EC}/m_e)}{I_0 (Q_{EC}/m_e)}  
\qquad ,
\qquad
I_k  (x_0)= \int_{1}^{x_0}  x^{1 - k} \, (x_0 - x)^2 \, \sqrt{x^2 - 1} \ dx~,
\ee
where $Q_{EC} = M_1 - M_2$ with $m_e$ the electron mass. 
The transition strengths, or $ft$ values, after the application of 
transition-dependent radiative corrections and 
isospin-symmetry-breaking corrections to the nuclear matrix elements, 
are remarkably constant with the 
$Z$ of the daughter nucleus,   
supporting the Conserved-Vector-Current (CVC)  ``hypothesis", 
 though CVC is simply a consequence of the SM. Moreover, CVC is 
also tested by studies of nuclear mirror transitions, albeit
with lesser precision~\cite{NaviliatCuncic:2008xt}. 
%
%
%
The computation of the isospin-symmetry-breaking corrections, 
noting e.g. Ref.~\cite{Hardy2004id}, has been 
criticized~\cite{Miller:2008my,Miller:2009cg}, but the employed procedure has been 
experimentally validated in a system for which the correction is particularly 
large~\cite{Melconian:2011kk}. 
For a discussion of the various methods to compute isospin-breaking effects, see Ref.~\cite{Towner:2010bx}.
From the constancy of the 
corrected $ft$ values with the $Z$ of the daughter nucleus, 
Hardy and Towner~\cite{Hardy:2008gy} obtain the combined bound 
\be
 - 1.0 \times 10^{-3}  <   g_S \,{\rm Re}\, \epsilon_S  <  3.2 \times 10^{-3}
 \qquad \qquad (90 \% \  \rm{C.L.}) \,.
\label{eq:bFbound}
\ee
It is the most stringent bound on scalar interactions from low-energy probes.
Moreover, the $\Delta_{\rm CKM}$ constraint implies
\be
{\rm Re} (  \epsilon_L + \epsilon_R  - \epsilon_\mu )   <  5  \times 10^{-4}
 \qquad \qquad (90 \% \  \rm{C.L.})~.
\ee
This is one of the strongest precision constraints on new physics.
It corresponds to effective scales $\Lambda > 11$~TeV~\cite{Cirigliano:2009wk}.
The resulting constraints on 
the weak-scale couplings 
$\hat{\alpha}_j$~\cite{Cirigliano:2009wk, Bhattacharya:2011qm,Cirigliano:2012ab}
are comparable to that obtained from Z-pole experiments,
and are stronger than the ones obtained from $\sigma(e^+ e^- \to q \bar{q})$ at LEP.

In principle, neutron decay allows for an extraction of ${\overline V}_{ud}$ 
free of nuclear structure
uncertainties.
Assuming no BSM effects one has 
$V_{ud} = 
[ 4908.7 (1.9) s/  ( \tau_n ( 1 + 3 g_A^2)]^{1/2}$~\cite{Czarnecki:2004cw,Marciano:2005ec}.
An extraction of $\overline{V}_{ud}$ competitive with nuclear decays requires
$\delta g_A/g_A \sim 0.025\%$
and $\delta \tau_n \sim 0.35 s$ ($\delta \tau_n/\tau_n = 0.04\%$).  In turn,  a determination of $g_A$ at the required
level necessitates a measurement of the beta asymmetry $A$ at the $0.1\%$ level.
The expression for 
$|\overline{V}_{ud}|^2 \Big|_{n \to p e \bar{\nu}}$  reads~\cite{Bhattacharya:2011qm}
\be
|\overline{V}_{ud}|^2 \Big|_{n \to p e \bar{\nu}}
= |V_{ud}|^2  \Bigg[  1
+  2 \,{\rm Re}  ( \epsilon_L  +  \epsilon_R  - \epsilon_\mu)
+
 \frac{ 2 }{1 + 3 \lambda^2}  \Bigg( g_S \,{\rm Re}\,  
\epsilon_S  -   12 \lambda  \, g_T \,{\rm Re}\, \epsilon_T  \Bigg)
 \left( \frac{I_1(x_0)}{I_0(x_0)} -  \frac{6 \lambda^2}{1+3\lambda^2}c \right) \Bigg]~ \,,
\label{eq:Vudn}
\ee
where $\lambda \equiv g_A/g_V$, $x_0 = E_0/m_e$, with $E_0$ the electron endpoint energy.
The constant $c$ is a certain ${\cal O}(1)$ 
number that depends on the specific experimental analysis
used to extract  $\lambda$ from measurements of the beta asymmetry $A$, presuming
the presence  of spectrum 
contaminations due to scalar and tensor operators.
A different correction would appear if one extracted  $\lambda$ from a different observable, such as $a$ 
(see the discussion in Ref.~\cite{Bhattacharya:2011qm}).

\subsection{\it Lepton Universality}
\label{sect:LFU}

The ratio $R_\pi \equiv \Gamma ( \pi \to e \nu [\gamma]) /
\Gamma ( \pi \to \mu \nu [\gamma])$
is  helicity-suppressed in the SM,
due to the $V-A$ structure of charged current couplings.
It is therefore a  sensitive probe of  all  SM extensions
that induce  axial and especially pseudoscalar currents, as well as of 
non-universal corrections to the  charged current lepton couplings.
The quantity 
$R_\pi$  can be predicted very precisely in the SM
because the leading hadronic  input, namely, the
pion decay constant $F_\pi$, cancels in the ratio.
Once one includes electroweak radiative corrections, hadronic structure effects do appear, 
and the SM prediction can be organized 
within the ChPT power counting as follows:
\bea
R_\pi^{\rm SM} &= & R_\pi^{(0)}
\, \Bigg[   1 +
\Delta_{e^2 p^2}  +   \Delta_{e^2 p^4}  +  ...
\Bigg]  \ \ \ \  \\
R_\pi^{(0)}  & =& \frac{m_e^2}{m_\mu^2}  \left(  \frac{m_\pi^2 - m_e^2}{m_\pi^2 - m_\mu^2}
 \right)^2  ~.
\label{eq:R0}
\eea
The leading electromagnetic correction  $\Delta_{e^2 p^2}$
 corresponds to  the point-like approximation for the pion~\cite{Kinoshita:1959ha,Marciano:1993sh}.
The NNLO (two-loop) correction  $\Delta_{e^2 p^4}$ has been calculated within ChPT in Refs.~\cite{Cirigliano:2007xi,Cirigliano:2007ga}.
The two-loop effective theory results have been complemented by a large-$N_c$
calculation of an associated counterterm and by 
summation of leading logarithms $\alpha^n \ln^n (m_\mu/m_e)$~\cite{Marciano:1993sh}
giving~\cite{Cirigliano:2007xi,Cirigliano:2007ga}
\be
R_{\pi} = (1.2352 \pm 0.0001) \times 10^{-4}~,
\ee
The central value of  $R_\pi$
is in agreement with the results of
previous calculations~\cite{Marciano:1993sh,Finkemeier:1995gi}, pushing the theoretical
uncertainty below the $0.1$ per-mille level.

\subsubsection{\it Model independent constraints}

The ratio $R_\pi \equiv \Gamma ( \pi \to e \nu [\gamma]) /
\Gamma ( \pi \to \mu \nu [\gamma])$
probes more than the effective low-energy pseudoscalar coupling
$\epsilon_P$ defined earlier as the coefficient of the operator
$\bar{e}(1 - \gamma_5) \nu_e \cdot \bar{u} \gamma_5 d$.  In fact,
since (i) $R_\pi$ is defined as the ratio of electron-to-muon decay
and (ii) the neutrino flavor is not observed in either decay, this
observable is sensitive to the whole set of parameters
$\epsilon_P^{\alpha \beta}$ 
and $\tilde{\epsilon}_P^{\alpha \beta}$  
defined by
\be
{\cal L}_{\rm eff} \quad  \supset    \quad
\frac{G_F}{\sqrt{2}} V_{ud} 
\Bigg[
\ \epsilon_P^{\alpha \beta} \  \ \bar{e}_\alpha (1 - \gamma_5) \nu_\beta \cdot \bar{u} \gamma_5 d \ + \  
\tilde{ \epsilon}_P^{\alpha \beta} \  \ \bar{e}_\alpha (1+  \gamma_5) \nu_\beta \cdot \bar{u} \gamma_5 d 
\Bigg]
\ee
where $\alpha \in \{e, \mu \}$ refers to the flavor of the charged lepton and  $\beta \in \{ e, \mu, \tau \}$
refers to the neutrino flavor.
One generically expects SM extensions to generate non-diagonal components in $\epsilon_{P,S,T}^{\alpha \beta}$. 
In the new notation the previously defined pseudoscalar, scalar, and tensor  couplings
read  $\epsilon_{P,S,T}  \equiv \epsilon_{P,S,T}^{ee}$.
It is important to note that only $\epsilon_P^{ee}$ and
$\epsilon_P^{\mu \mu}$ can interfere with the SM amplitudes, while the remaining
$\epsilon_P^{\alpha \beta}$ 
and $\tilde{\epsilon}_P^{\alpha \beta}$  
each enter as an absolute square in 
the numerator and denominator of $R_\pi$.
In summary, allowing for non-standard  axial and pseudoscalar interactions and factoring out the SM prediction for $R_\pi$,
one can write:~\cite{Bhattacharya:2011qm} 
\be
\label{eq:Rpiconstraint1}
\frac{R_\pi}{R_\pi^{\rm SM}} =
\frac{\left[ \left\vert 1 + \epsilon_L^{ee} - \epsilon_R^{ee}  - \frac{B_0}{m_e} \epsilon_P^{ee} \right\vert^2
+  \left\vert\frac{B_0}{m_e} {\epsilon}_P^{e\mu} \right\vert^2
+  \left\vert\frac{B_0}{m_e}{\epsilon}_P^{e\tau} \right\vert^2
+  \sum_\alpha  \left\vert\frac{B_0}{m_e}{\tilde{\epsilon}}_P^{e\alpha} \right\vert^2
 \right]}{
\left[ \left\vert 1 + \epsilon_L^{\mu \mu}- \epsilon_R^{\mu \mu} - \frac{B_0}{m_\mu} \epsilon_P^{\mu \mu} \right\vert^2
+  \left\vert\frac{B_0}{m_\mu}{\epsilon}_P^{\mu e} \right\vert^2
+  \left\vert\frac{B_0}{m_\mu}{\epsilon}_P^{\mu \tau} \right\vert^2
+  \sum_\alpha  \left\vert\frac{B_0}{m_\mu}{\tilde{\epsilon}}_P^{\mu \alpha} \right\vert^2
\right]} \ \equiv \ 1 + \Delta_{e/\mu}~\,,
\ee
where we note that the BSM couplings can be complex. 
In the above equation the factors of $B_0/m_{e, \mu} \epsilon_P$ represent the ratio
of the new-physics amplitude over the SM amplitude.  The latter is proportional to
the charged-lepton mass due to angular-momentum conservation arguments, while the former
is proportional to  $\langle 0| \bar{u} \gamma_5 d| \pi\rangle$, characterized by
the  scale- and scheme-dependent parameter\footnote{Note that the scale and scheme dependence
of $B_0(\mu)$ is compensated in physical quantities  by the scale  and 
scheme dependence of the Wilson coefficients $\epsilon_P^{\alpha \beta}$.}
\be
B_0 (\mu)   \equiv  \frac{M_\pi^2}{m_u (\mu) + m_d(\mu)}~.
\ee
Since $B_0^{\overline{\rm MS}} (\mu = 1 \  {\rm GeV}) = 1.85 \, {\rm GeV}$ 
(using the PDG~\cite{Nakamura:2010zzi}  central values for the light quark masses)  and consequently
$B_0/m_e =  3.62 \times 10^3$,  $R_\pi$ has enhanced sensitivity to $\epsilon_P^{\alpha \beta}$, and
one needs to keep quadratic terms in these new physics coefficients. 
We discuss bounds on 
$\epsilon_A \equiv \epsilon_L - \epsilon_R$ 
and $\epsilon_P^{\alpha \beta},  \tilde{\epsilon}_P^{\alpha \beta}$  separately. 
First, setting $ \epsilon_A = 0$, 
inspection of  Eq.~(\ref{eq:Rpiconstraint1}) reveals that
 if the new-physics couplings respect 
 $\epsilon_P^{e \alpha}/m_e =   \epsilon_P^{\mu  \alpha}/m_\mu$,
 then $R_\pi/R_\pi^{\rm SM} =1$, and there are no constraints on these couplings.
On the other hand, if the effective  couplings $\epsilon_P^{\alpha \beta}$ are all of similar size,
one can neglect the entire denominator in Eq.~(\ref{eq:Rpiconstraint1}),
as it is suppressed with respect to the numerator by powers of $m_e/m_\mu$.
We will assume the second scenario operates.
In this case the constraint in
Eq.~(\ref{eq:Rpiconstraint1}) forces the couplings
$\epsilon_P^{ee},\epsilon_P^{e\mu},\epsilon_P^{e \tau}, \tilde{\epsilon}_P^{e \alpha}$ to live in a spherical shell
of radius   $m_e/B_0 \sqrt{R_\pi^{\rm exp}/R_\pi^{\rm SM}} \approx  2.75 \times 10^{-4}$  centered at
${\rm Re} (\epsilon_P^{ee}) = m_e/B_0 \approx 2.75 \times 10^{-4}$, $\epsilon_P^{e\mu}= \epsilon_P^{e \tau}= \tilde{\epsilon}_P^{e \alpha} = 0$.
The thickness of the shell is numerically $1.38 \times 10^{-6}$
and  is determined by the current combined uncertainty
in  $R_\pi^{\rm exp}$~\cite{Britton:1992pg,Britton:1993cj,Czapek:1993kc}
and $R_\pi^{\rm SM}$~\cite{Cirigliano:2007ga,Cirigliano:2007xi}:
$R_\pi^{\rm exp}/R_\pi^{\rm SM} = 0.996(5)$ (90\% C.L.).
This is illustrated in Fig.~\ref{fig:Remu}, where we plot the allowed region
in the two-dimensional plane given by ${\rm Re} ( \epsilon_P^{ee}) $ and a generic
``wrong-flavor" coupling denoted by $\epsilon_P^{ex}$ ($x \neq e$)
--- or this can be ${\rm Im} (\epsilon_P^{ee})$  or  any of the real or imaginary parts of 
$\epsilon_P^{e x}$  and  $\tilde{\epsilon}_P^{e \alpha}$. 
Note that the allowed region is given by   the
thickness of the curve in the figure,
thus enforcing a strong correlation between
 $\epsilon_P^{ee}$ and $\epsilon_P^{ex}$.
Since $\epsilon_P^{e x}$ and others of that sort 
are essentially unconstrained by other measurements, 
though we expect they can be of ${\cal O}(10^{-3})$, 
we can 
neglect all of the couplings but one to obtain a bound on that one. 
The resulting bounds using $R_\pi$ at  90\%-C.L. are 
\be
\label{eq:epsPconstraint}
- 1.4  \times 10^{-7}  <  {\rm Re} \left(  \epsilon_P^{ee}   \right) <  5.5 \times 10^{-4},  \quad  {\rm or} \quad
- 2.75  \times 10^{-4}  <  {\rm Im} \left(  \epsilon_P^{ee} \right)   <  2.75 \times 10^{-4} \,,
\ee
Note that 
${\rm Re} (\epsilon_P^{e x})$ and ${\rm Im} (\epsilon_P^{e x})$, 
as well as ${\rm Re}  (\tilde{\epsilon}_P^{e \alpha })$ and 
${\rm Im} (\tilde{\epsilon}_P^{e \alpha })$, 
are all subject to the same bound as  ${\rm Im} (\epsilon_P^{ee})$. 
Our results are 
in qualitative agreement with the findings of  Refs.~\cite{Herczeg:1994ur,Herczeg2001vk}.

Alternatively, 
neglecting the pseudoscalar couplings, one obtains the following combined limit 
on the axial combination of new couplings: 
\be
- 4.5  \times 10^{-3}  <   {\rm Re} \left( \epsilon_A^{ee} - \epsilon_A^{\mu \mu}  \right)  <  0.5 \times 10^{-3}    \quad ~.
\ee

\begin{figure}[t!]
\centering
\includegraphics[width=0.50\textwidth]{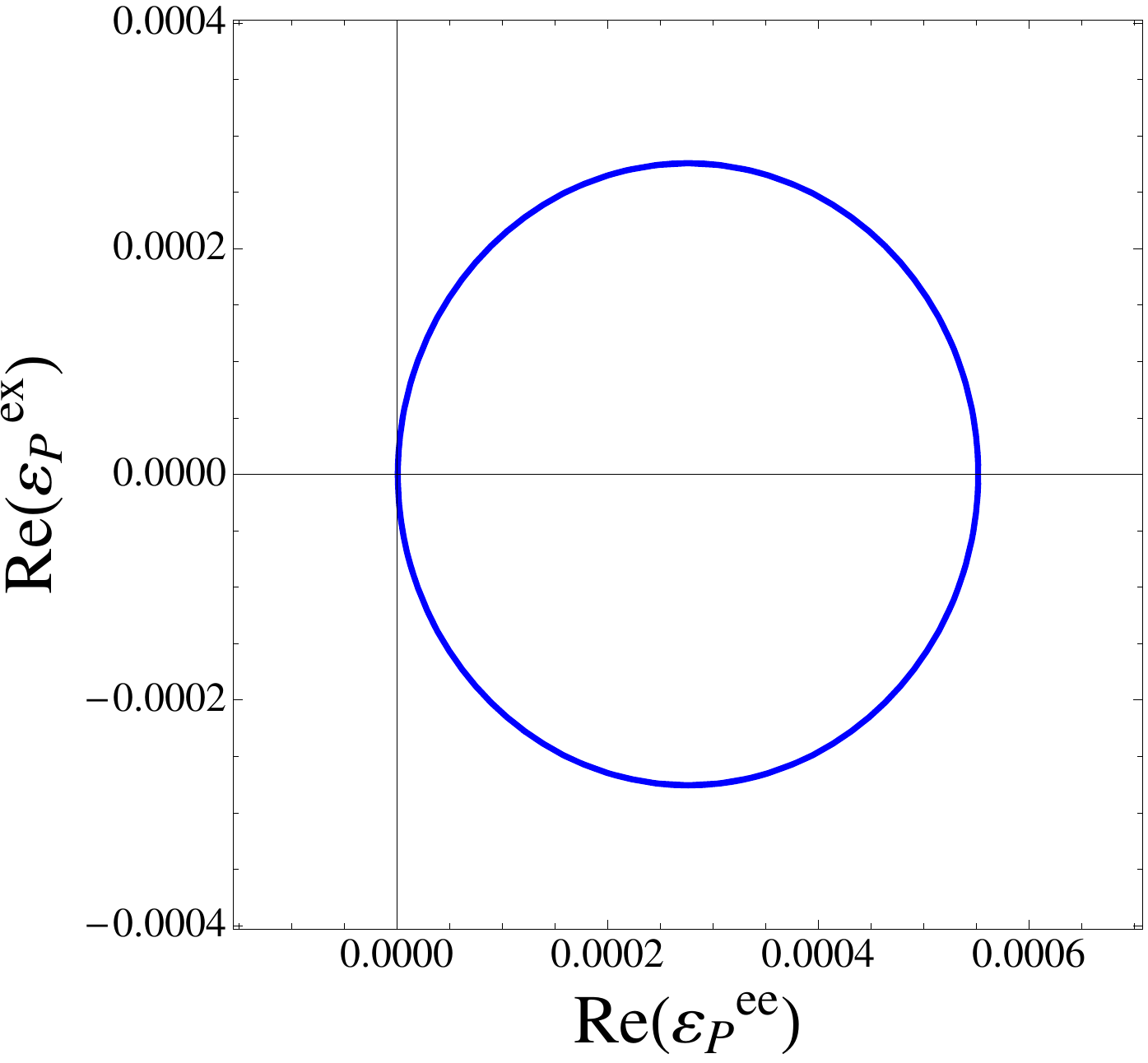}
\begin{minipage}[t]{16.5 cm}
\caption{Illustration of 
the allowed region in the two-dimensional plane  ${\rm Re} (\epsilon_P^{ee})$-${\rm Re} (\epsilon_P^{ex})$
(with $x \neq e$) determined by
$R_\pi$, which is given by  an annulus of thickness $1.38 \times 10^{-6}$.
In the  absence of information on ${\rm Re} (\epsilon_P^{ex})$,
the 90 \% C.L. bound on ${\rm Re} (\epsilon_P^{ee})$ is
$- 1.4  \times 10^{-7}  <  {\rm Re} (\epsilon_P^{ee} ) <  5.5 \times 10^{-4}$.
Note that  $ {\rm Im} (\epsilon_P^{e \alpha})$,
 ${\rm Re}  (\tilde{\epsilon}_P^{e \alpha })$, and 
${\rm Im} (\tilde{\epsilon}_P^{e \alpha})$ 
are subject to the same bound as  ${\rm Re} (\epsilon_P^{e x})$. 
Figure adapted  from Ref.~\cite{Bhattacharya:2011qm}.
}
\label{fig:Remu}
\end{minipage}
\end{figure}

Finally, we discuss how $R_\pi$ is also sensitive to non-standard scalar and tensor couplings.
As originally discussed in Refs.~\cite{Voloshin:1992sn,Herczeg:1994ur, Campbell:2003ir},
the pseudoscalar coupling $\epsilon_P^{ee}$ can be radiatively generated
starting from nonzero $\epsilon_{S,T}$. Hence, the stringent constraint in Eq.~(\ref{eq:epsPconstraint})
puts constraints on the same $\epsilon_{S,T}$ that can be probed in beta decays.
The physics of the effect is this: 
once the scalar, pseudoscalar, and tensor operators are generated
by some non-standard physics at the matching scale $\Lambda$, 
electroweak radiative corrections
induce mixing among these three operators. Thus even if 
$\epsilon_P (\Lambda)$ vanishes at the matching scale,  known SM physics
generates a nonzero $\epsilon_P (\mu)$ at some lower energy  scale $\mu$  via loop diagrams.
The general form of the constraint can be worked out by using the
three-operator mixing results from Ref.~\cite{Campbell:2003ir}.
The leading-order result is
\begin{subequations}
\bea
\epsilon_P^{\alpha \beta} (\mu) &=& \epsilon_P^{\alpha \beta}  (\Lambda) \left(1 + \gamma_{PP} \,  \log \frac{\Lambda}{\mu} \right)
+ \epsilon_S^{\alpha \beta}  (\Lambda)  \ \gamma_{SP} \,  \log \frac{\Lambda}{\mu}
+ \epsilon_T^{\alpha \beta}  (\Lambda)\  \gamma_{TP} \,  \log \frac{\Lambda}{\mu}
\\
\gamma_{PP} &=& \frac{3}{4} \frac{\alpha_2}{\pi}  + \frac{113}{72} \frac{\alpha_1}{\pi}  \approx 1.3 \times 10^{-2}
\\
\gamma_{SP} &=&  \frac{15}{72} \frac{\alpha_1}{\pi}  \approx 6.7 \times 10^{-4}
\\
\gamma_{TP} &=& -  \frac{9}{2} \frac{\alpha_2}{\pi}  -  \frac{15}{2} \frac{\alpha_1}{\pi} \approx - 7.3 \times 10^{-2}~,
\eea
\end{subequations}
where $\alpha_1 = \alpha/\cos^2\theta_W$ and $\alpha_2 = \alpha/\sin^2\theta_W$ are the U(1) 
and SU(2)$_L$
weak couplings, expressed in terms of the fine-structure constant and the weak mixing angle.
Setting $\epsilon_P^{ee} (\Lambda) = 0$  and neglecting the small 
${\cal O}(\alpha/\pi)$ fractional difference between
$\epsilon_{S,T} (\Lambda)$ and the observable
$\epsilon_{S,T} (\mu)$ at the low scale,
the constraints  on $\epsilon_S$ and-$\epsilon_T$ using
$R_\pi$ at 90\% C.L. read 
\be
\frac{- 1.4 \times 10^{-7} }{\log (\Lambda/\mu)}
 \ <  \  \gamma_{SP}  \  {\rm Re} ( \epsilon_S )  \ + \   \gamma_{TP}  \  {\rm Re} ( \epsilon_T )    \ <  \ \frac{5.5 \times 10^{-4}}{\log (\Lambda/\mu)}~, 
\ee
and 
\be
| \gamma_{SP}  \  {\rm Im } ( \epsilon_S )  \ + \   \gamma_{TP}  \  {\rm Im } ( \epsilon_T )    | \ <  \ \frac{2.75 \times 10^{-4}}{\log (\Lambda/\mu)}~, 
\ee
Assuming $\log (\Lambda/\mu) \sim 10$, so that, e.g., 
$\Lambda \sim 10\, {\rm TeV}$ and $\mu \sim  1 \, {\rm GeV}$, and 
using the numerical values of $\gamma_{SP,TP}$, 
one finds  that the individual constraints are at the  level of
$| {\rm Re} (\epsilon_S)  |  \lsim  8 \times 10^{-2}$,
$|{\rm Im } (\epsilon_S)  | \lsim  4 \times 10^{-2}$,
$| {\rm Re} ( \epsilon_T )|   \lsim  10^{-3}$, and 
$| {\rm Im } ( \epsilon_T )|   \lsim 0.5 \cdot 10^{-3}$.
These bounds become logarithmically more stringent as the new-physics scale $\Lambda$ grows.
It is worth noting that analogous studies are also possible in kaon decays, and new
results are expected from NA62 at CERN~\cite{NA62:2011aa}  and TREK at J-PARC~\cite{Shimizu:2012zzb}. 

Constraints  on $\tilde{\epsilon}_{S,T}$ can be worked  out 
similarly~\cite{Campbell:2003ir}, resulting in 
 $| {\rm Re} (\teS )|  \lesssim  5 \times 10^{-2}$, 
 $| {\rm Im } (\teS )|  \lesssim  2.5 \times 10^{-2}$, 
 $|{ \rm Re}  (\teT )|  \lesssim 0.6 \times 10^{-3}$, 
and 
 $|{ \rm Im} ( \teT) |  \lesssim 0.3 \times 10^{-3}$, 
  which together with
$\tilde{\epsilon}_P$ are the strongest  low-energy 
bounds on the $\tilde{\epsilon}$ couplings~\cite{Cirigliano:2012ab}.

\section{Decay  correlations and  non-$(V-A)$ couplings}
\label{sect:correlations}

Differential decay distributions in beta decays
are very sensitive to the Lorentz structure of the underlying weak interaction,
thus enabling searches for small non-$(V-A)$  components.
Following Ref.~\cite{Jackson1957zz}, one writes 
the differential decay distribution 
 in the nuclear decay  $P  \to D  \ e^{-} \ \bar{\nu} \ (e^+ \nu) $
as a function of electron (positron) energy and lepton directions 
as follows,
\be
\frac{d^3 \Gamma}{dE_e d\Omega_e d\Omega_\nu} =   \frac{1}{(2\pi)^5}p_e E_e (E_0 - E_e)^2 \xi
\left\{1  
+  b \, \frac{m_e}{E_e}   +
   a \, \frac{\vec{p}_e \cdot \vec{p}_\nu}{E_e E_\nu}   
+ 
 \langle \frac{\vec{J}}{J} \rangle \cdot  \left[ A \, \frac{\vec{p}_e}{E_e}  +   B \, \frac{\vec{p}_\nu}{E_\nu}
 +
 D \, \frac{\vec{p}_e \times \vec{p}_\nu}{E_e \, E_\nu}  \right]
 + \ldots 
 \right\} \,,
 \label{eq:correlations}
\ee
where $P$ and $D$ represent the parent and daughter nuclei,
$\langle \vec{J} \rangle/J$ represents the parent nucleus polarization, if any,  and
$\vec{p}_{e,\nu}$ are the electron (positron) and antineutrino (neutrino) three-momenta.
We have omitted the additional parity-conserving term which appears if $J\ne 1/2$, 
as indicated by the ellipsis.
The coefficient $b$ is the Fierz interference term, 
$a$ is the electron-antineutrino correlation, $A$ is the beta asymmetry,
$B$ is the antineutrino asymmetry, and the coefficient $D$ 
is $T$ odd in that the associated triple product of vectors is motion-reversal odd. 
All these quantities contain combinations of the Lee-Yang effective 
coefficients as delineated in Ref.~\cite{Jackson1957zz}, 
as does $\xi$, and are related to our parameters as per Eq.~(\ref{eq:conversions}).
Additional terms are present 
if one can measure the polarization of the emitted electron or positron~\cite{Jackson1957zz}.
Note, too, that the various correlation coefficients become 
$E_e$ dependent once corrections of radiative and recoil order are included. 

The decay correlations can be measured in neutron and nuclear decays,
and substantial progress is expected in the next few years.
In neutron decay, both cold and ultracold neutrons, implying distinct experimental 
techniques and hence entirely independent sources of systematic error, are used 
to measure
these correlations. In the future we can expect experiments poised to 
take advantage of cold neutron beams of much greater intensity at the 
FRM-III (PERC)~\cite{Dubbers:2011ns}, the New Guide Hall at 
NIST~\cite{nico}, and the SNS (Nab)~\cite{abBA}. 
Concerning nuclear decays, the development of 
atomic trapping techniques has allowed the precise detection of daughter
nucleus recoil momenta, which in turn 
permits bettered measurements of the electron-antineutrino correlation parameter $a$.

In the absence of radiative corrections, recoil corrections, and BSM contributions,
the correlation coefficients $a (E_e)$, $A (E_e)$, and $B (E_e)$ 
reduce to simple expressions, while $b$,$D=0$ vanish.
For example, for a pure Gamow-Teller decay
we have the prediction $a_{GT}=-1/3$,
whereas for a pure Fermi transition we have $a_F=1$. 
For mixed Fermi-Gamow-Teller transitions there is also a precise prediction 
once the ratio of Fermi to Gamow-Teller strengths is known---and this can be
determined from the lifetime. 
In the case of neutron decay, which is a mixed transition, one obtains:
\bea
{a}(E_e) \to  \frac{1 - \lambda^2}{1 + 3 \lambda^2} ~,
~~~~~~{A}(E_e) \to    \frac{2 \lambda (1 - \lambda)}{1 + 3 \lambda^2} ~,
~~~~~~{B}(E_e) \to     \frac{2 \lambda ( 1 + \lambda)}{1 + 3 \lambda^2}~,
\eea
where $\lambda \equiv g_A/g_V$ and the limiting value of $B(E_e)$, e.g., 
is termed $B_0$. 

Going beyond the SM, the dependence of the correlations $a,b,A,B$, and $D$ 
on the short-distance couplings
$\epsilon_i$ and $\tilde{\epsilon}_i$ can be determined  using their dependence on
the couplings $C_i \pm C_i'$  given in Ref.~\cite{Jackson1957zz}
and the relations given in Eq.~(\ref{eq:conversions}).
The full expressions are quite complicated, but simplify considerably 
if one considers the leading {\it linear}
corrections only. In regards to these, the salient points are: 

\begin{itemize}
\item  As mentioned previously, 
the right-handed coupling  $\epsilon_{R}$  to linear order induces
  the shift $\lambda \to \tilde{\lambda} = \lambda (1 - 2 \epsilon_R)$.
 In order to probe $\epsilon_R$ from correlation measurements, 
one needs to know $\lambda \equiv g_A/g_V$ independently; 
this can come from a LQCD calculation.

\item The scalar and tensor couplings $\epsilon_{S,T}$
appear at linear order only  through the Fierz interference term ${b}$ and the analogous term
$b_\nu$ in the antineutrino asymmetry parameter, where $b_\nu$ is defined by
$B(E_e) = B_0  + b_\nu  m_e/E_e$.
Different nuclear transitions probe different combinations of the BSM couplings.
For example, the Fierz term $b$ in a pure Fermi or Gamow-Teller transition probes
exclusively the scalar or tensor coupling, according to
$b_{\rm F} = \mp  2 \gamma \, g_S \, {\rm Re} (\epsilon_S)$
and
$b_{\rm GT}= \pm (8 \gamma  g_T {\rm Re} ( \epsilon_T)) /\lambda$, 
where $ \gamma = \sqrt{ 1- \alpha^2  Z^2}$ 
and the sign distinguishes $\beta^{\pm}$ emitters~\cite{Jackson1957206}. 
Mixed transitions such as neutron decay probe  a linear combination of
scalar and tensor couplings. For neutron decay one has~\cite{Jackson1957206}:
\begin{subequations}
\label{eq:bbsm}
\bea
b &=&  \frac{ 2 \gamma }{1 + 3 \lambda^2}  \Bigg[ g_S \, {\rm Re} 
  (\epsilon_S )  -   12 \lambda  \, g_T \, {\rm Re}  (  \epsilon_T )  \Bigg]
~,
\\
b_\nu &=&  \frac{2\gamma}{1 + 3 \lambda^2}  \Bigg[ g_S \, {\rm Re}  ( \epsilon_S ) \,  \lambda   - 4 g_T \,
{\rm Re}  ( \epsilon_T)   \, (1 + 2 \lambda)  \Bigg]
~.
\eea
\end{subequations}

\item Measurements of the  correlation coefficients  $a,A$, and $B$
always include contributions from the Fierz interference term  $b$,
and are therefore sensitive to $\epsilon_{S,T}$ to linear order.
This dependence arises because
correlation measurements involve the construction of asymmetry ratios~\cite{Gluck:1995hs}, 
and the dependence on $b$ does not cancel in the asymmetry denominators.
For example, in order to isolate ${A} (E_e)$ one constructs the ratio
\be
A_{\rm exp} (E_e) = \frac{N_+ (E_e) - N_- (E_e)}{N_+ (E_e) + N_- (E_e)} \,,
\ee
where $N_{\pm} (E_e)$ are the spectra corresponding to events with
$\vec{J} \cdot  {\vec p} _e >0$ and
$\vec{J} \cdot  {\vec p} _e <0$, respectively, so that 
sensitivity to  ${b}$ does indeed appear through the denominator.
In general, asymmetry measurements probe 
\be
\tilde{Y} (E_e)  = \frac{{Y} (E_e)}{ 1 + {b}  \, m_e/E_e + \ldots}~,  \qquad \qquad  Y \in \{ A,B, a, ... \} \,,
\label{eq:obstilde}
\ee
where the ellipsis denotes other possible corrections of radiative and recoil order whose
appearence depend on the correlation considered. It is 
worth noting that simultaneous analysis of $a(E_e)$ and $A(E_e)$, e.g., yields more powerful
constraints on the underlying BSM contributions than either correlation alone~\cite{svgbp}.

\end{itemize}
The dependence on  $\tilde{\epsilon}_\alpha$ couplings appears only to quadratic order,
together with additional effects quadratic in the $\epsilon_\beta$'s. In this section, 
we have considered beta-decay correlations only, though strong constraints on BSM
couplings also come from the study of light meson decays. 
We now review constraints on the various BSM couplings which appear in 
Eq.~(\ref{eq:conversions}).

\subsection{\it Model-independent constraints on scalar and tensor couplings}

We now summarize the  current best constraints on the scalar and tensor
structures,  and highlight  prospects for future improvements.
Currently,  the most stringent constraint
on the scalar coupling $\epsilon_S$  arises from $0^+ \to 0^+$
nuclear beta decays, as discussed in Section~\ref{sec:cabibbo1}, whereas 
the most stringent bound
on the tensor effective coupling $\epsilon_T$ arises from the Dalitz-plot
study of the radiative pion decay $\pi^+  \to  e^+ \nu_e \gamma$.
That is, an analysis of the Dalitz plot of  this decay  from the PIBETA
collaboration~\cite{Bychkov:2008ws}
puts constraints on the product ${\rm Re} (\epsilon_T f_T)$ 
of the short-distance coupling $\epsilon_T$
and the hadronic form factor $f_T$  defined by~\cite{Mateu:2007tr}
\be
\langle \gamma (\epsilon, p) | \bar{u} \sigma_{\mu \nu} \gamma_5 d | \pi^+  \rangle =
- \frac{e}{2}  \, f_T  \, \left(p_\mu \epsilon_\nu - p_\nu \epsilon_\mu  \right)~,
\ee
where $p_\mu$ and $\epsilon_\mu$ are the photon four-momentum and 
polarization vector, respectively.
The analysis of Ref.~\cite{Mateu:2007tr}, based on a 
large-$N_c$-inspired resonance-saturation model, 
provides  $f_T = 0.24(4)$ at the renormalization scale $\mu=1$~GeV,
with the parametric uncertainty induced by the uncertainty in the quark condensate.
The $90\%$-C.L. experimental constraint\footnote{Note 
that there is a factor of 2 difference in the
normalization of the tensor coupling $\epsilon_T$ compared to what was used in
Refs.~\cite{Herczeg:1994ur,Bychkov:2008ws}.}
$ - 2.0 \times 10^{-4}  \ < \  f_T  {\rm Re} ( \epsilon_T )   <  \   2.6  \times 10^{-4} $,  when combined with
the  above estimate for $f_T$ evolved to 2~GeV implies
\be
- 1.1  \times 10^{-3}  \ < \ {\rm Re} ( \epsilon_T ) \  <  \  1.36  \times 10^{-3} \qquad \qquad (90 \%  \ \rm{C.L.})~.
\label{eq:tradpi}
\ee
This is the most stringent constraint on the tensor coupling from low-energy experiments.
The next best constraints  arise from measurements of nuclear beta decays~\cite{Severijns2006dr}.

Bounds on scalar and tensor interactions can be obtained from a number of observables
in nuclear beta decays, other than $0^+ \to 0^+$ transitions.
Although these bounds are currently not competitive, we summarize them here for completeness.
The leading sensitivity to scalar and tensor operators appears through the Fierz interference term $b$.
Significant constraints on $b$ arise from the 
electron-polarization observables~\cite{Jackson1957zz}
as well as from measurements of  ${A}$  and ${a}$
in Fermi, Gamow-Teller, and mixed transitions.
Here is a summary of  current  bounds on $\epsilon_{S,T}$~\cite{Severijns2006dr}:
\begin{itemize}
\item The most stringent constraint from the beta asymmetry in pure Gamow-Teller transitions
($\tilde{A}_{\rm GT}$)  arises from  $^{60} {\rm Co}$ 
measurements and implies~\cite{Wauters:2010gh}
\be
- 2.9 \times 10^{-3}  \ < \ g_T \, 
{\rm Re}
 (\epsilon_T)  \  <  \   1.5  \times 10^{-2} \qquad \qquad (90 \%  \ \rm{C.L.})~.
\ee
Similar bounds can be obtained from  
measurements of $\tilde{A}_{\rm GT}$ in $^{114}{\rm In}$ 
decay~\cite{Wauters:2009jw}:  
$- 2.2 \times 10^{-2}  \ < \ g_T \,
{\rm Re}
( \epsilon_T ) \  <  \  1.3 \times 10^{-2}$ (90 \% C.L.).

\item Measurements of the ratio $P_{\rm F}/P_{\rm GT}$ from the 
longitudinal polarization of the positron emitted in pure Fermi and Gamow-Teller
transitions~\cite{Carnoy:1991jd,Wichers:1986es} imply
\be
- 0.76  \times 10^{-2}  \ < \    g_S \, {\rm Re} (\epsilon_S )  +
    \frac{4}{\lambda}   g_T \, {\rm Re} ( \epsilon_T ) \  <  \   1.0  \times 10^{-2} \qquad \qquad (90 \%  \ \rm{C.L.})~.
\ee

\item    Preliminary results have been reported on the 
measurement of the longitudinal polarization of positrons emitted by
polarized $^{107} {\rm In}$ nuclei~\cite{Severijns:2000vc}. 
The corresponding  90 \% C.L. sensitivity to tensor interactions,
 $|   g_T \, {\rm Re}  ( \epsilon_T ) |   <    3.1 \times 10^{-3}$, is quite promising 
although not yet competitive with radiative pion decay.

\item Finally,  the beta-neutrino correlation $a$ has been measured in a number of nuclear
transitions~\cite{Vetter:2008zz,Gorelov:2004hv,Adelberger:1999ud,Johnson:1963zza}.
The resulting constraints on scalar and tensor interactions are summarized in
Fig.~7 of Ref.~\cite{Vetter:2008zz}.
In terms of the coupling constants used here, the 90 \% C.L. combined 
bound on the tensor interaction reads  $| g_T \,
{\rm Re} ( \epsilon_T)  | <  5  \times 10^{-3}$, 
again not competitive with radiative pion decay.

\end{itemize}

Future improvements can be expected from both neutron and nuclear decay measurements.
In the case of neutrons~\cite{Abele:2008zz,Dubbers:2011ns},
future measurements of 
the beta asymmetry $A$~\cite{PERKEOIII:2009,Plaster:2008si,abBA,Dubbers:2007st}, 
the antineutrino asymmetry $B$~\cite{WilburnUCNB,abBA},
the electron-neutrino correlation $a$~\cite{Pocanic:2008pu,aSPECT:2008,Wietfeldt:2005wz},
and the Fierz interference term $b$~\cite{Pocanic:2008pu,UCNb}
should exceed $10^{-3}$ precision.
On the nuclear side, measurements of $a$ and $b$ in the pure Gamow-Teller decay of
$^6$He~\cite{Knecht201143} should also reach the $10^{-3}$ level in precision.
In Fig.~\ref{fig:epsST} we summarize the current constraints on $\epsilon_S$ and $\epsilon_T$
(horizontal bands) and assess the impact of future measurements 
in both neutron and nuclear  decays, where we assume  $|b,b_\nu|< 10^{-3}$
from neutron decay  and $|b_{\rm GT}| < 10^{-3}$ from nuclear decays  at 90\% CL.
The future neutron constraints are represented by the diagonal bands, whereas the
constraint from $^6$He is represented by the vertical ocher band in the plot. 
In  Fig.~\ref{fig:epsST}  the left panel represents the constraints 
using quark model input for the scalar and tensor matrix elements $g_{S,T}$, 
whereas the right panel uses our preferred LQCD estimates~\cite{Bhattacharya:2011qm}, 
which still have a 50\% uncertainty in $g_S$.  In the near future one
can expect $\delta g_S /g_S$ to reach the 20\% level from LQCD, thus increasing
the constraining power of these measurements, so that the thickness of the bands will shrink.
With current uncertainties on $g_{S,T}$, measurements of $10^{-3}$-level precision 
will probe effective scales $\Lambda_{S,T} > 7$~TeV.

\begin{figure}[t]
\centering
\includegraphics[width=0.9\textwidth]{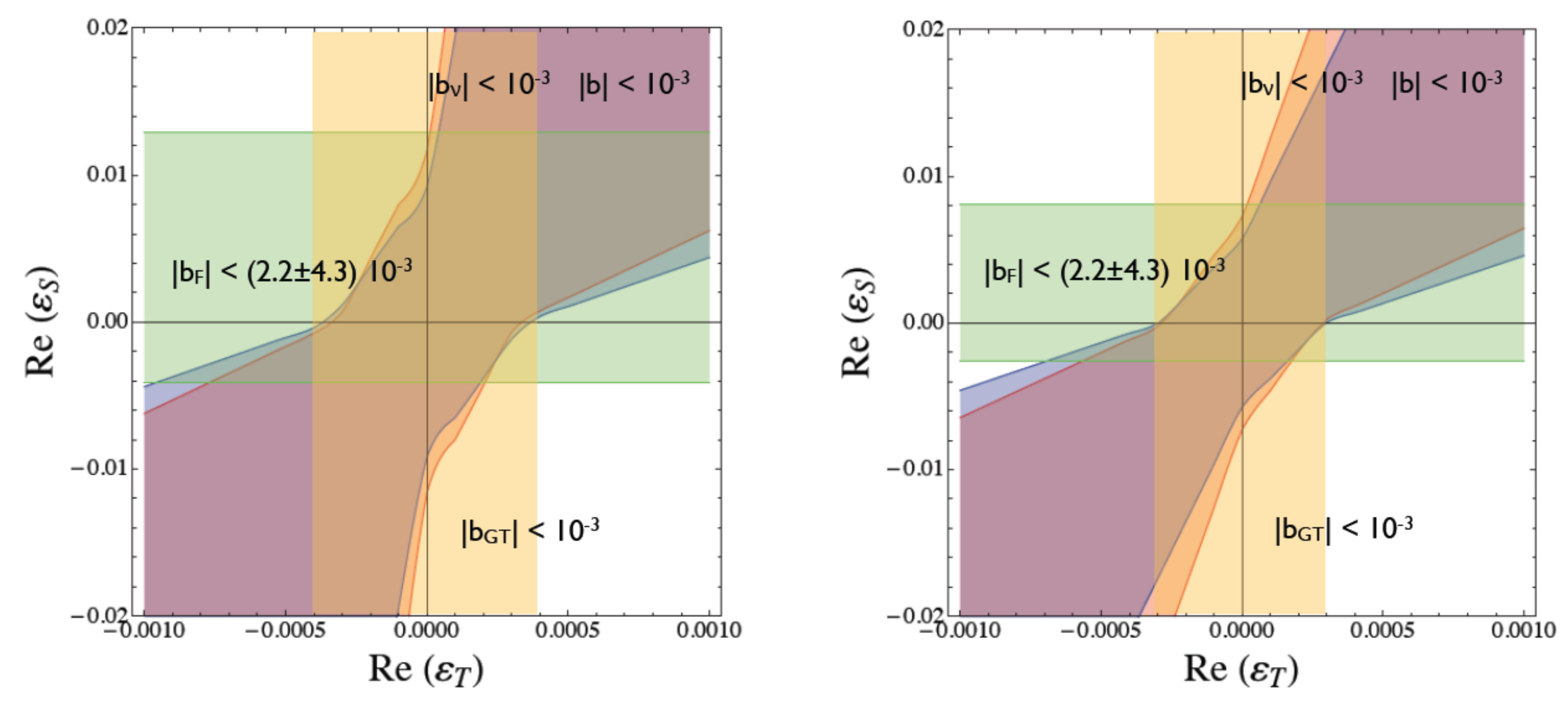}
\begin{minipage}[t]{16.5 cm}
\caption{
Current and prospective 
$90\%$ C.L. allowed regions  in
the  ${\rm Re} (\epsilon_S)$-${\rm Re} (\epsilon_T)$ plane
implied by (i)  the existing bounds on $b_{\rm F}$
and $\pi \to e \nu \gamma$
(horizontal (green) band);
(ii)  projected measurements of $b$ and $b_\nu$ in neutron decay
(inner (red)  and  outer (blue) bow-tie shaped regions, respectively)
 at the $10^{-3}$ level;
(iii) projected measurements of $b_{\rm GT}$ at the $10^{-3}$ level
from $^6$He decays (vertical (ocher) band).
Left panel:    hadronic matrix elements taken in the  ranges
$0.25 < g_S < 1.0$,   $0.6 < g_T < 2.3$~\cite{Herczeg2001vk}.
Right panel:  scalar and tensor charges taken from
LQCD,  $g_{S}= 0.8(4)$ and  $g_{T}= 1.05(35)$.
Note that by reducing the uncertainty in $g_S$
the constraint on $\epsilon_S$  from $b_{\rm F}$  becomes stronger,
independent of any future neutron measurement.
The effective couplings $\epsilon_{S,T}$ are defined 
in the $\overline{\rm MS}$ scheme at 2~GeV. Figure adapted from 
Ref.~\cite{Bhattacharya:2011qm}.}
\label{fig:epsST}
\end{minipage}
\end{figure}

\subsection{\it Nuclear and neutron probes of recoil effects}

SM recoil effects can be tested to high accuracy via
experiments  involving correlations between final state electron/positrons and the
polarization or alignment of the parent state, which have the form
\be
{d^3\Gamma\over dE_ed\Omega_ed\Omega_\nu}\sim 1+A(E_e)  \, P  \,  {\hat{z}\cdot\vec{p}_e\over E_e}+F(E_e)\Lambda\left({1\over E_e^2}\hat{z}\cdot\vec{p}_e\hat{z}\cdot\vec{p}_e-{p_e^2\over 3E_e^2}\right)
\ee
where $P=<J_z>/ J$ is the polarization 
and $\Lambda=1-3<J_z^2>/(J(J+1))$ is the alignment. 
The beta correlation $A$ receives both leading order and recoil corrections, 
whereas the alignment correlation $F$ is purely a recoil order effect. 
However, both have been measured as functions of $E_e$ 
and provide measures of the recoil form factors. 
In this regard, it is important to point out that 
the axial tensor form factor $d(q^2)$, which vanishes in the SM 
from isotopic spin invariance in the case of transitions 
between isotopic analog states such as in neutron beta decay or tritium decay, 
is in general {\it nonvanishing}. Indeed, it is in general 
comparable in size to the weak magnetism term $\tilde{b}(q^2)$\footnote{Note 
that the vanishing of $d(0)$ for transitions between isotopic analog states 
is a SM prediction and is violated by so-called second class 
currents which can arise if quarks have an additional 
quantum number\cite{Holstein:1976mw}.}.  In the 
 case of mirror transitions the size of the axial tensor 
must be identical for electron and positron decays, and this is subject to test. 
A theoretical prediction for the weak magnetism term $\tilde{b}$ in terms 
of the difference between parent and daughter magnetic moments exists for 
transitions between isotopic analogs, and in the case of mirror 
transitions the size of the weak magnetism term is given 
by the electromagnetic M1 width of the transition 
from the excited isotopic analog state of the daughter 
nucleus---another CVC test~\cite{GellMann:1958zz}. One expects corrections to this
result which are linear in isospin-breaking, both in $\tilde b(0)$ itself and from 
possible second-class current contribution in the SM. 
In neutron beta decay, through study of $a(E_e)$ and $A(E_e)$, 
the second-class current contribution to $d(0)$ 
can be determined independently of $\tilde b(0)$, implying that the 
CVC test can be made without assumptions regarding second-class 
currents~\cite{Gardner:2000nk}. 
Generally, since the form of the recoil effects to all observables has been given, 
together with electromagnetic corrections, these 
SM predictions are all testable~\cite{Holstein:1974zf}.

An alternate route for testing the SM 
recoil predictions is to utilize decays where the daughter state is unstable 
and itself decays
\begin{itemize}
\item [i)] electromagnetically such as in the mirror transitions 
in the $A=20$ system to the $2^+$ 1.63 MeV excited state of $^{20}$Ne, 
which in turn decays via photon emission to the $0^+$ ground state~\cite{Minamisono:2011zz};
\item [ii)]  strongly such as in the mirror transitions 
in the $A=8$ system to the $2^+$ 2.90 MeV excited state of $^8$Be, 
which in turn decays via the emission of two alpha particles~\cite{Sumikama:2011pr}.
\end{itemize}
In the former case there exists a beta-gamma correlation
\begin{equation}
{d^3\Gamma\over dE_ed\Omega_ed\Omega_\gamma}\sim 1+{1\over 2}G(E_e)\left(\left({\vec{p}_e\cdot\hat{p}_\gamma\over E_e}\right)^2
-{p_e^2\over 3E_e^2}\right)\,,
\end{equation}
whereas in the latter there is a beta-alpha correlation
\begin{equation}
{d^3\Gamma\over dE_ed\Omega_ed\Omega_\alpha}\sim 1+G(E_e)\left(\left({\vec{p}_e\cdot\hat{p}_\alpha\over E_e}\right)^2
-{p_e^2\over 3E_e^2}\right)-2{\vec{p}_e\cdot\hat{p}_\alpha\over Mv*}\,,
\end{equation}
noting $v*$ is the velocity of the alpha particle in the daughter rest frame.  
Here the decay correlation coefficient $G$ is purely of recoil order, so that it 
is sensitive to recoil form factors. 
The form of $G$ as well as of the radiative corrections have 
been calculated~\cite{Holstein:1974zf}.

\subsection{\it  Couplings involving right-handed neutrinos:   $\tilde{\epsilon}_{L,R,S,T,P}$}

Neglecting neutrino masses, all the $\tilde{\epsilon}_\beta$ couplings enter 
the  decay rates quadratically, i.e., as per $\propto | \tilde{\epsilon}_\beta|^2$.
Detailed expressions of the contributions to 
neutron and nuclear beta decay correlation coefficients
 can be found  in Ref.~\cite{Jackson1957206}, though 
one needs to re-express the Lee-Yang couplings in 
terms of the  $\epsilon_\alpha$ and  $\tilde{\epsilon}_\beta$ 
using  Eq.~(\ref{eq:conversions}). 
The corresponding bounds can be obtained from the analysis of  Ref.~\cite{Severijns2006dr},
in particular from the fits to beta decay data allowing for non-zero
$\teL, \teR$ only,
implying $| {\rm Re} (  \teL + \teR)| < 6.4 \times 10^{-2}$,  $|
{\rm Re} (\teL -  \teR| )  < 5.8 \times 10^{-2}$,
and  $\teS, \teT$ only, implying $|g_S    {\rm Re} ( \teS ) | < 5.5 \times 10^{-2}$,
$|g_T  {\rm Re} (\teT )| <  2.1  \times 10^{-2}$ at 90\% CL.
Decay correlations set also bounds on the imaginary parts of these couplings. 
For example, using the measurements of $a$ in the decay $^{32}$Ar~\cite{Adelberger:1999ud}  
and $^6$He~\cite{PhysRev.132.1149,Gluck1998493}, one gets 
$|{\rm Im}(\tilde{\epsilon}_S)| < 0.17$ and 
$ |{\rm Im} (\tilde{\epsilon}_T)| < 0.05$ at 90\% CL. 
We expect that these  bounds can  be improved 
by future more precise measurements.

\subsection{\it T-odd correlations}

Triple-product decay correlations can only be motion-reversal-odd 
and thus cannot be true tests of T-invariance~\cite{Sachs:1987gp}. 
As a consequence final-state interactions (FSI) can simulate 
a seeming T-odd correlation without a fundamental
violation of T-invariance. In beta-decays the energy release is 
sufficiently small that the FSI are 
electromagnetic in nature, so that they are calculable with minimal hadronic ambiguity
at accessible levels of precision~\cite{Callan:1967zz,Okun:1967ww,Ando:2009jk}. 
Nevertheless, under an assumption of CPT invariance,
such ``T-odd'' correlations are sensitive to new  sources of CP violation,
and thus to new physics, though an observation of such a correlation in excess
of SM expectations would not allow one to conclude 
that T itself is violated~\cite{Wolfenstein:1999re}. 

T-odd correlations have been studied in kaon, neutron,  hyperon~\cite{hyperT,Marchioro:1969ds}, and nuclear 
beta-decays. 
In $K^+$ decay, namely, $K^+ \to \mu^+ \nu \gamma$, 
the transverse muon polarization is studied, and the expected SM correlation
is small~\cite{Hiller:1999xa,Braguta:2002gz} with respect to that possible
in models of new physics~\cite{Wu:1996hi,Hiller:1999xa}. Existing experimental
studies are consistent with no T-violation in this and related 
processes~\cite{Anisimovsky:2003me,Abe:2006de}, but
new results of greater sensitivity 
are expected from TREK at J-PARC~\cite{Kohl:2010zz,Djalali:2012zz}. 
In ordinary beta-decay, a T-odd correlation is possible only if
the initial state is polarized, or if the final-state polarization of one
of the particles is observed. 
In the decay of polarized neutrons one can construct
$\vec{J} \cdot (\vec{p}_e \times \vec{p}_\nu)$, i.e., the $D$ correlation, 
or 
$\vec{J} \cdot (\vec{p}_e \times \vec{\sigma}_e)$, i.e., the $R$ correlation, 
if the polarization of the emitted electron $\propto \vec{\sigma}_e$ is observed.
Recently significant experimental efforts in regard to each of these correlations
have been concluded. 
The  emiT collaboration has presented the best limit on $D$ in beta 
decay~\cite{Chupp:2012ta,Mumm:2011nd}, finding 
$D_n = (-0.94 \pm 1.89 \pm 0.97) \times 10^{-4}$ in neutron decay, 
a substantial improvement over
earlier measurements~\cite{Lising:2000pa,Soldner:2004xm,Severijns2006dr}. 
As for $D_{FSI}$, the ${\cal O}(\alpha)$ 
correction vanishes in the zero-recoil limit, 
and $D_{FSI} \approx 10^{-5}$~\cite{Callan:1967zz,Holstein:1983cv}. This calculation
has been updated to employ the techniques of heavy-baryon chiral 
effective field theory by Ando et al.~\cite{Ando:2009jk}; 
 they reproduce the Callan-Treiman result in 
${\cal O}(\alpha \bar Q/M_N)$ with $\bar Q\sim M_n - M_p - m_e$ and include the 
leading piece of the N${}^3$LO correction to realize $D_{FSI}$ with an estimated accuracy
of better than 1\%. 
In terms of our non-standard couplings 
$D$ in neutron decay can be written as~\cite{Jackson1957zz}
\begin{equation}
D_{BSM} = \frac{1}{1 + 3 \lambda^2}  \ \Bigg[ 4 \lambda \, {\rm Im} (\epsilon_R)   \ + \ 
8 g_S g_T   \, {\rm Im} \left(   \epsilon_S \epsilon_T^*  \, + \, \tilde{\epsilon}_S \tilde{\epsilon}_T^* \right)~ . 
\Bigg]
\end{equation}

Neglecting small quadratic effects in  scalar and tensor couplings, 
the emiT limit translates  to  the 90\% C.L.  constraint    $- 5 \times 10^{-4} <   {  \rm Im} ( \epsilon_R ) <    3 \times 10^{-4}$,  
where we have used $\tilde\lambda$~\cite{Beringer:2012zz}
for $\lambda$. 
For  ${\rm Im} (\epsilon_{S,T})$ and  ${\rm Im} (\tilde{\epsilon}_{S,T})$ 
the analysis is more involved. 
$D$  provides bounds on products such as  ${\rm Re} (\epsilon_T) {\rm Im} (\epsilon_S)$, etc.,  
so that no bounds on the imaginary parts can be 
obtained unless non-zero real parts of the exotic couplings 
are assumed or discovered. 

Recent results also exist for $R$. In the first neutron experiment 
both transverse components of the 
electron polarization are measured, to yield both $R$ and $N$~\cite{Kozela:2009ni}. 
The correlation $N$ 
probes $\vec{J}\cdot \vec{\sigma}_e$ and is appreciably non-zero from FSI; 
the experimental measurement is consistent with the SM expectation, offering
a sensitivity check of the setup~\cite{Kozela:2011mc}. They find 
$R_n= 0.004\pm 0.012 \pm 0.005$, 
limiting  the imaginary parts of scalar and tensor interactions via: 
\be
R_n = \frac{1}{1 + 3 \, \lambda^2}  \, \Big[
- 8   g_T    \left( 2 \lambda  + 1  \right) {\rm Im} (\epsilon_T) 
 \ - \ 2 g_S \, \lambda \, {\rm Im} (\epsilon_S)
 \Big]~.
\ee 
In comparison the measurement
of $R$ in $^8$Li decay yields $R_{^8{\rm Li}}=0.0009(22)$~\cite{Huber:2003gr},
which through 
\be
R_{^8{\rm Li}} =  - \frac{1}{3} \frac{8 g_T}{g_A}  \, {\rm Im} (\epsilon_T) 
\ee
implies a limit 
$- 3.1 \times 10^{-3}  <  {\rm Im} (\epsilon_T)   <  1.8  \times 10^{-3} $ at 90\% C.L.
after $R_{FSI}$ has been removed. 
Consequently, the $R_n$ limit is most usefully interpreted as a limit on 
${\rm Im} (\epsilon_S)$, namely 
$- 0.15 <  {\rm Im} (\epsilon_S)   < 0.11$ at 90\% C.L.
Looser  constraints  (at the 20\%-level) on the imaginary
parts of the scalar and tensor couplings come from 
the measurement of $a$ in $0^+\to 0^+$ 
transitions~\cite{Adelberger:1999ud,Gorelov:2004hv} 
and in ${}^6$He decay~\cite{Johnson:1963zza}, respectively---note Fig.~22 of 
Ref.~\cite{Chupp:2012ta} for a useful compilation.

The possibility of  constraining T-odd P-even couplings, including $\beta$ decay parameters,  
via T-odd P-odd observables such as EDMs has been considered earlier in Ref.~\cite{Khriplovich:1990ef}. 
The basic idea is that through loop diagrams involving  electroweak gauge boson exchanges, 
T-odd P-even interactions can generate T-odd P-odd interactions. 
The resulting bounds depend on the scale at which parity invariance is restored~\cite{RamseyMusolf:1999nk,Kurylov:2000ub}.
In this context, the $D$ correlation has been recently studied in Ref.~\cite{Ng:2011ui},
considering how large $D$ 
can be in light of constraints from electric dipole moment (EDM) searches.  
Focusing on  the leading contribution to $D$ (proportional to 
${\rm Im} (\epsilon_R)$)  the authors show that via a loop diagram 
the same phase contributes to the neutron  and other EDMs.
If  this is the leading (or only) contribution to the 
neutron EDM, then one can conclude that  
the neutron EDM currently provides 
the strongest constraint on $D$, which is $10 - 10^3$
times stronger than current direct limits on $D$, depending on the model. 
Of course, the bounds on $D$ can be weakened if other operators,  
other than the 4-fermion ones, 
contribute to the neutron EDM, and this is natural in 
many theories, 
 interfering destructively with the 
contribution proportional to ${\rm Im} (\epsilon_R)$. 
The numerical evaluation of the effect of such 
operators becomes a model-dependent question,   
and one anticipates that the connection between $d$ to $D$ can be 
weakened completely, 
though detailed investigation is warranted~\cite{svgdD}.

In radiative $\beta$ decay one can form a T-odd correlation from momenta alone, so
that one probes new spin-independent sources of CP violation. A triple momentum
correlation has been previously studied in 
$K^+ \to \pi^0 l^+ \nu_l \gamma$~\cite{Braguta:2001nz}, and its sensitivity to 
physics BSM considered~\cite{Braguta:2003wf}. 
In $K^+$ decay both electromagnetic
and strong, i.e., pion-mediated, radiative corrections can mimic the T-odd effect, but 
the electromagnetic-induced FSI are orders 
of magnitude larger~\cite{Braguta:2001nz,Khriplovich:2010rz,Muller:2006gu}. 
Finally,  a spin-independent 
T-odd correlation can be constructed in the 
radiative beta decay of neutron and nuclei~\cite{Gardner:2012rp},
proportional to $\vec{p}_\gamma \cdot( \vec{p}_e \times \vec{p}_\nu)$, offering 
the opportunity to study the imaginary part of the pseudo-Chern-Simons 
term~\cite{Gardner:2013aiw} first found as a consequence of the baryon vector current anomaly and 
SU(2)$_L\times$U(1) gauge invariance at 
low energies~\cite{Harvey:2007rd,Harvey:2007ca,Hill:2009ek}. 
In Ref.~\cite{Gardner:2012rp}, the effect of FSI 
on this new correlation have been computed,
establishing the baseline for possible future searches of BSM CP-violating interactions.

\begin{figure}[t]
\begin{center}
\includegraphics[width=0.50\hsize]{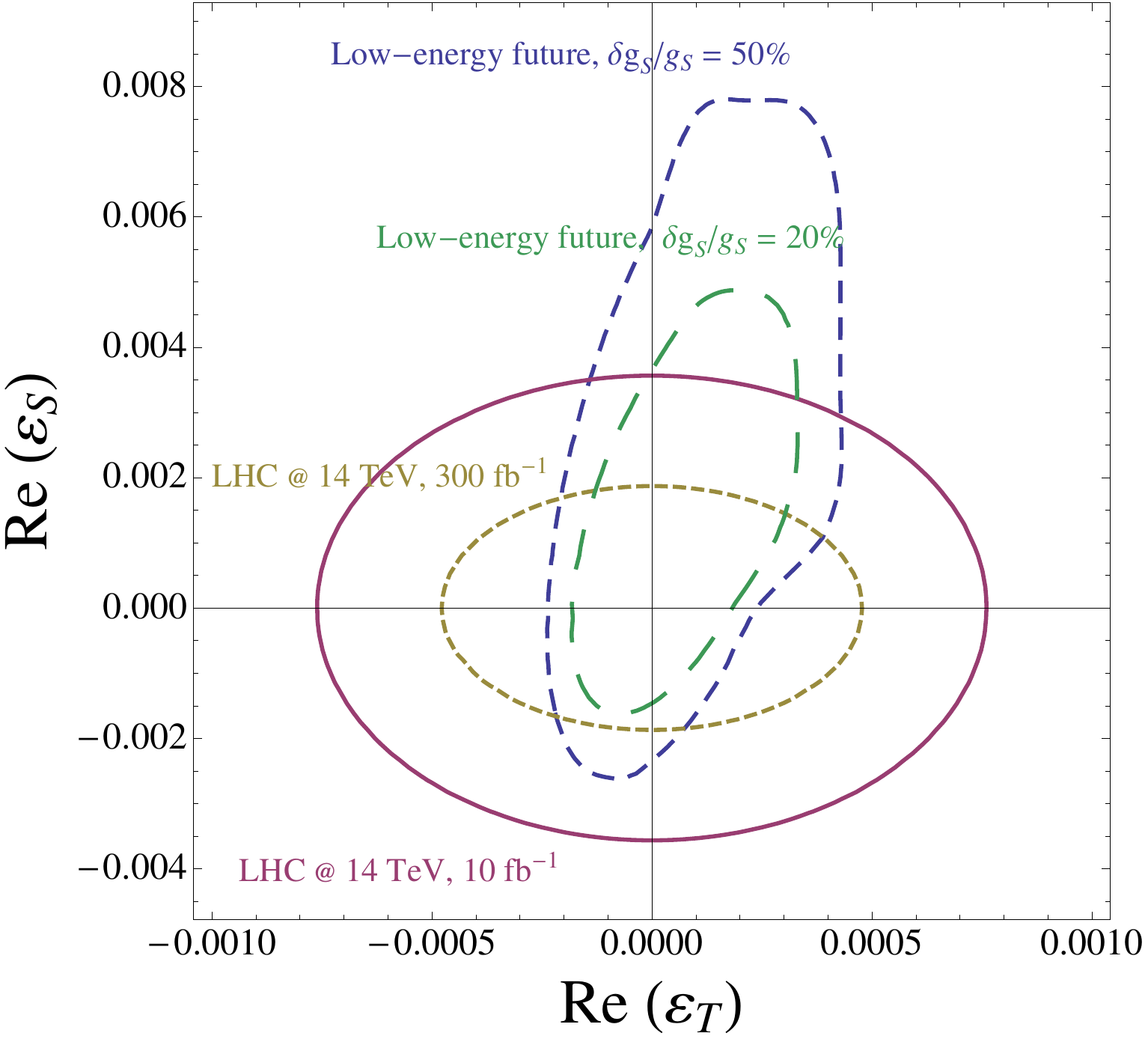}
\begin{minipage}[t]{16.5 cm}
\caption{\small  
Projected joint $90 \%$ CL constraints  on ${\rm Re} (\epsilon_S)$ and ${\rm Re} (\epsilon_T)$ from 
future  beta decays measurements and the LHC at $\sqrt{s}=14$ TeV.
The low-energy constraints correspond to 0.1\% measurements of $B,b$ in neutron decay and 
$b$ in $^6$He decay, under two different scenarios for the lattice QCD uncertainties in $g_{S,T}$.
The LHC bounds are obtained  by  requiring less than 3 $e \nu$-produced signal events with: (i) $m_T>$ 2.5 TeV and
$10$ fb$^{-1}$ of integrated luminosity (solid, red ellipse); and (ii) $m_T>4$ TeV and 300 fb$^{-1}$ (dashed, yellow ellipse).
Cuts are chosen to reduce the expected leading background to be below 1 event. To obtain the projection it is assumed no events are found.
Note that the effective couplings  $\epsilon_{S,T}$ are defined  in the $\overline{\rm MS}$ scheme at 2~GeV.
Figure adapted  from Ref.~\cite{Bhattacharya:2011qm}.
\label{LHCboundb}}
\end{minipage}
\end{center}
\end{figure}

\section{Collider limits on non-standard CC interactions}
\label{sect:collider}

The BSM  interactions probed at low energy can also be directly probed at high-energy colliders.
The collider signals, however, depend
on whether the particles that generate the 4-fermion interactions are kinematically accessible at the collider energies.
Model-independent statements can be made
under the assumption that the non-standard interactions remain point-like 
at TeV-scale energies; i.e., the mediators
are not accessible at the LHC. In this case
collider searches directly  probe the various non-standard 
couplings $\epsilon_\alpha$  and $\tilde{\epsilon}_\beta$, 
that contribute to the parton-level amplitude for  $pp  \to  e \nu + X$. 
Since the non-standard amplitudes do not interfere with the SM amplitude (except for terms proportional to $\epsilon_L$), 
 the LHC probes at the same level both the real and imaginary parts of the new couplings.  
In Ref.~\cite{Bhattacharya:2011qm} bounds are 
derived on $\epsilon_{S,T}$ by analyzing LHC data in the   $pp  \to  e \nu + X$ channel
at $\sqrt{s} =7$~TeV and 1 fb$^{-1}$ integrated luminosity.
In Ref.~\cite{Cirigliano:2012ab}  the analysis has been extended to all 
non-standard charged-current couplings with integrated luminosity of 5 fb$^{-1}$, 
using both  the $pp  \to  e \nu + X$ and $pp  \to e^+ e^-  + X$  channels 
through SU(2)$_L$ gauge invariance.

The   $pp  \to  e \nu + X$ channel  is directly related to 
beta decays, since the parton-level process is $\bar{u} d \to  e \bar{\nu}$.
In order to put bounds on the BSM couplings one uses
the (cumulative) transverse mass distribution~\cite{ENcms5fb}, noting that 
the transverse mass of the lepton pair 
is defined as $m_{T} \equiv \sqrt{2 E^e_T E^\nu_T (1 - \cos \Delta \phi_{e\nu}}$.
At high $m_T$ the SM background falls off while the 
BSM interactions would produce events and thus increase their number. 
Similarly,  for $pp  \to e^+ e^-  + X$ one uses 
the dilepton invariant mass distribution~\cite{EEcms5fb}, dubbed 
$m_{ee}$,  to constrain the presence of possible contact interactions.

A comparison of the best bounds available for each interaction from low- and high-energy
experiments is shown in   Tables \ref{tab:summaryLRE}, \ref{tab:summaryLIM} 
(for ${\rm Re} (\epsilon_\alpha) $  and ${\rm Im} (\epsilon_\alpha)$)  
and  \ref{tab:summaryRRE},   \ref{tab:summaryRIM} 
(for ${\rm Re} (\tilde{\epsilon}_\alpha)$  and ${\rm Im} (\tilde{\epsilon}_\alpha)$). 
Note that in these tables we report only direct bounds, leaving out bounds on the 
real and imaginary parts of $\epsilon_{S,T}$ and $\tilde{\epsilon}_{S,T}$ from $R_\pi$ 
as they can be evaded by cancellation. 
All of the tabulated results refer to a bound on the absolute value
of the parameter unless a range is specified. 
The main points can be summarized as follows~\cite{Cirigliano:2012ab} 
(see also Ref.~\cite{martinoscar}). 
For the pseudo-scalar couplings $\eP$ and $\teP$ 
the low-energy constraints from pion decay are at the $10^{-4}$ level, which are 
very hard to reach at the LHC in the near future.  The same applies to the 
vector interactions $\epsilon_{L,R}$, 
for which both the CKM-unitarity bound (${\rm Re} (\epsilon_{L,R})$) and 
the emiT bound  (${\rm Im} (\epsilon_{R})$)  are at the $0.5  \times 10^{-3}$-level. 

For scalar and tensor interactions 
with left-handed neutrinos, 
low-energy experiments have traditionally yielded stronger bounds
on ${\rm Re} (\eS) $ and ${\rm Re} (\eT)$,  
but the current LHC bounds have caught up; and 
both probes are at the $10^{-2}$ and $10^{-3}$ level for 
${\rm Re} (\eS) $ and ${\rm Re} (\eT)$,   respectively.
In the next few years we expect improvements in the bounds from both the LHC
and  low-energy experiments, through  neutron~\cite{Dubbers:2007st,abBA,WilburnUCNB,Pocanic:2008pu,PERKEOIII:2009,UCNb}
and $^6$He decay~\cite{Knecht201143} measurements at the 0.1\% level and 
beyond.
Projected future bounds from both beta decays at the $10^{-3}$ level and the 
LHC on ${\rm  Re}  (\epsilon_{S,T})$ are shown in  Figure~\ref{LHCboundb}.
These results show that low-energy searches with $10^{-4}$ sensitivity would have
unmatched constraining potential, even in the LHC era.
Concerning the imaginary parts, for ${\rm Im} (\epsilon_T)$ 
bounds from low-energy ($R$ correlation in $^8$He) and the LHC are at the same level, 
while  for ${\rm Im} (\epsilon_S)$  the LHC bound is stronger than the one derived from the 
$R$ correlation  in neutron decay.

Finally, for scalar and tensor interactions with right-handed 
neutrinos, $\teS$ and $\teT$, the LHC bounds are 
also at the $10^{-2}$ and $10^{-3}$ level respectively, 
significantly better than the current and future
low-energy limits 
for  both   ${\rm Re} (\tilde{\epsilon}_{S,T}) $ and ${\rm Im} (\tilde{\epsilon}_{S,T})$. 
To match the LHC bound one needs measurements of the
 electron-neutrino correlation $a$ in Gamow-Teller 
transitions at the level of 
$\delta a_{GT}/a_{GT} \sim 0.05$~\%.\footnote{Within a given NP model the ratio 
$\Gamma(\pi\to e\nu) / \Gamma(\pi\to \mu\nu)$ is 
likely to produce the strongest bound not only on
$\epsilon_P$ and $\teP$,  but also on
$\epsilon_{S,T}$ and $\tilde{\epsilon}_{S,T}$, 
through their loop-induced contribution. However, since
these bounds are based on  naturalness arguments, they could be circumvented by
 cancellations between different contributions,  so that 
in a model-independent analysis the LHC offers 
the strongest constraint on $\tilde{\epsilon}_{S,T}$.}.
Similar remarks apply to $\teR$,  for which the
LHC bound is $5\times 10^{-3}$, 
and no significant limit is available from low-energy probes.

\begin{table}[h]
\begin{center}
\begin{minipage}[t]{16.5 cm}
\caption{Summary of 90\% CL bounds (in units of  $10^{-2}$)
on the real parts of  non-standard couplings $\epsilon_\alpha$  obtained  from low-energy and LHC searches 
(5~fb$^{-1}$  at $\sqrt{s} = 7$~TeV).
In order to deduce the low-energy bounds on the scalar  and tensor couplings
we used  $g_S = 0.8(4)$ and $g_T = 1.05(35)$~\cite{Bhattacharya:2011qm}.
Using  $g_S = 1.08(28)$~\cite{Green:2012ej} 
would lead to the stronger bound $| {\rm Re} (\epsilon_S)| < 0.4 \times 10^{-2}$.
The couplings  $\epsilon_{S,P,T}$    are evaluated  in the  $\overline{\rm MS}$  scheme at $\mu=2$~GeV.
\label{tab:summaryLRE}}
\end{minipage}
\begin{tabular}{c|c|c|c|c|c}
\hline
                     &       &            &            &              &                  \\[-2mm]
                        &	 $ {\rm Re}  ( \epsilon_L$ )     &       $  {\rm Re}  (\epsilon_R) $           &     $   {\rm Re} ( \epsilon_P )$    &      $ {\rm Re}  (\epsilon_S)$         &       ${\rm Re}  (\epsilon_T)$    \\
                     &       &            &            &              &                  \\[-2mm]
\hline
\hline
                     &       &            &            &              &                  \\[-2mm]
$\beta$
decays             &       0.05                  &               0.05                         &            0.04             &                   0.8                   &              0.1                 \\
                     &       &            &            &              &                  \\[-2mm]
\hline
                     &       &            &            &              &                  \\[-2mm]
 LHC
 $(e \nu)$        &   $(-0.3,+0.8)$    &                             --              &               1.3                   &                1.3                       &             0.3               \\
                     &       &            &            &              &                  \\[-2mm]
 \hline
                     &       &            &            &              &                  \\[-2mm]
  LHC
$(e^+ e^-)$   &		       		--      &    --                                      &           1.0                             &               1.0               &          0.1                 \\
                     &       &            &            &              &                  \\[-2mm]
\hline
\end{tabular}
\end{center}
\end{table}

\begin{table}[h]
\begin{center}
\begin{minipage}[t]{16.5 cm}
\caption{Summary of 90\% CL bounds (in units of  $10^{-2}$)
on the imaginary parts of  non-standard couplings $\epsilon_\alpha$  
obtained  from low-energy
(using hadronic input as per Table~\protect{\ref{tab:summaryLRE}}) 
and LHC searches (5~fb$^{-1}$  at $\sqrt{s} = 7$~TeV).
The couplings  $\epsilon_{S,P,T}$    are evaluated  in the  $\overline{\rm MS}$  scheme at $\mu=2$~GeV.
\label{tab:summaryLIM}}
\end{minipage}
\begin{tabular}{c|c|c|c|c|c}
\hline
                     &       &            &            &              &                  \\[-2mm]
                        &	 $ {\rm Im}  ( \epsilon_L$ )     &       $  {\rm Im}  (\epsilon_R) $           &     $   {\rm Im} ( \epsilon_P )$    &      $ {\rm Im}  (\epsilon_S)$         &       ${\rm Im}  (\epsilon_T)$    \\
                     &       &            &            &              &                  \\[-2mm]
\hline
\hline
                     &       &            &            &              &                  \\[-2mm]
$\beta$
decays             &       --                 &               $(-0.05,+0.03) $                         &            0.02             &                   (-15,+11)                   &               $(-0.3, + 0.2) $                 \\
                     &       &            &            &              &                  \\[-2mm]
\hline
                     &       &            &            &              &                  \\[-2mm]
 LHC
 $(e \nu)$        &   0.5    &                        --                   &               1.3                   &                1.3                       &             0.3               \\
                     &       &            &            &              &                  \\[-2mm]
 \hline
                     &       &            &            &              &                  \\[-2mm]
  LHC
$(e^+ e^-)$   &		     --  		      &              --                            &           1.0                             &               1.0               &          0.1                 \\
                     &       &            &            &              &                  \\[-2mm]
\hline
\end{tabular}
\end{center}
\end{table}

\begin{table}[h]
\begin{center}
\begin{minipage}[t]{16.5 cm}
\caption{Summary of 90\% CL bounds, in units of  $10^{-2}$, 
on the real parts of  the non-standard couplings $\tilde{\epsilon}_\alpha$  
obtained  from low-energy and LHC searches (5~fb$^{-1}$  at $\sqrt{s} = 7$~TeV).
In order to deduce the low-energy bounds on the scalar  and tensor couplings
we used  $g_S = 0.8(4)$ and $g_T = 1.05(35)$~\cite{Bhattacharya:2011qm}.
Using  $g_S = 1.08(28)$~\cite{Green:2012ej} 
would lead to the stronger bound $| {\rm Re} (\tilde{\epsilon}_S)| < 6.9 \times 10^{-2}$.
The couplings  $\tilde{\epsilon}_{S,P,T}$    are evaluated  in the  $\overline{\rm MS}$  scheme at $\mu=2$~GeV.
\label{tab:summaryRRE}}
\end{minipage}
\begin{tabular}{c|c|c|c|c|c}
\hline
                     &       &            &            &              &                  \\[-2mm]
                     &                      $  {\rm Re} (\teL)$           &        ${\rm Re}  (\teR)$           &                                 $ {\rm Re} ( \teP) $    &                 $ {\rm Re} ( \teS) $         &                   $ {\rm Re} ( \teT)$               \\
                     &       &            &            &              &                  \\[-2mm]
\hline
\hline
                     &       &            &            &              &                  \\[-2mm]
$\beta$
decays       &   		6	                   &        6                     &                   0.02                             &                    14                    &                 3.0                                  \\
                     &       &            &            &              &                  \\[-2mm]
\hline
                     &       &            &            &              &                  \\[-2mm]
 LHC
 $(e \nu)$   &   			--	         	&	0.5	             &          1.3                                        &                   1.3               &                              0.3                                  \\
                     &       &            &            &              &                  \\[-2mm]
\hline
\end{tabular}
\end{center}
\end{table}

\begin{table}[h]
\begin{center}
\begin{minipage}[t]{16.5 cm}
\caption{Summary of 90\% CL bounds, in units of  $10^{-2}$, 
on the imaginary parts of the non-standard couplings $\tilde{\epsilon}_\alpha$  
obtained  from low-energy
(using hadronic input as per Table~\protect{\ref{tab:summaryLRE}}) 
 and LHC searches (5~fb$^{-1}$  at $\sqrt{s} = 7$~TeV).
 The couplings  $\tilde{\epsilon}_{S,P,T}$    are evaluated  in the  $\overline{\rm MS}$  scheme at $\mu=2$~GeV.
\label{tab:summaryRIM}}
\end{minipage}
\begin{tabular}{c|c|c|c|c|c}
\hline
                     &       &            &            &              &                  \\[-2mm]
                     &                      $  {\rm Im} (\teL)$           &        ${\rm Im}  (\teR)$           &                                 $ {\rm Im} ( \teP) $    &                 $ {\rm Im} ( \teS) $         &                   $ {\rm Im} ( \teT)$               \\
                     &       &            &            &              &                  \\[-2mm]
\hline
\hline
                     &       &            &            &              &                  \\[-2mm]
$\beta$
decays       &   		--	                   &        --                     &                   0.02                             &                   17                    &                5.0                                 \\
                     &       &            &            &              &                  \\[-2mm]
\hline
                     &       &            &            &              &                  \\[-2mm]
 LHC
 $(e \nu)$   &   			--	         	&	0.5	             &          1.3                                        &                   1.3               &                              0.3                                  \\
                     &       &            &            &              &                  \\[-2mm]
\hline
\end{tabular}
\end{center}
\end{table}

\section{Model constraints}
\label{sect:models}

Our discussion has focused on model-independent constraints emergent from 
precise universality tests and
correlation coefficient measurements, 
in terms of the effective couplings $\epsilon_\alpha$ and $\tilde{\epsilon}_\beta$.
As we have mentioned, ultraviolet extensions of the SM will  generate  
these non-standard effective couplings at some level, 
which are then functions of the model parameters.
Therefore, the results we have presented can be used 
to constrain  the parameter space  of any SM extension.
Moreover,  within a given SM extension,  the low-energy effective couplings will
show peculiar dependencies on the underlying model parameters,
resulting in  correlations among low-energy beta decay signatures 
and other observables.
In this subsection we briefly illustrate these ideas
by  discussing non-standard CC interactions  within  the 
Left-Right Symmetric Model~\cite{Mohapatra:1979ia,Holstein:1977qn}  and  the MSSM, 
for which detailed studies can be found in
Refs.~\cite{Barbieri:1985ff,Hagiwara:1995fx,Kurylov:2001zx,RamseyMusolf:2007yb,Bauman:2012fx}.
For more extensive reviews of underlying models 
we refer to~\cite{Herczeg2001vk}.

\subsection{\it Left-Right Symmetric Model}

The Left-Right Symmetric Model~\cite{Mohapatra:1979ia}
is based on an extended gauge group SU(2)$_L \times$ SU(2)$_R \times$ U(1),
in which in addition to the SM gauge  assignments,
the right-handed fermions transform as doublets under  SU(2)$_R$.
After spontaneous symmetry breaking, the charged gauge bosons $W_L$ and $W_R$ mix
into  light SM-like  $W_1$, which is predominantly left-handed, 
and heavier $W_2$, which is predominantly right-handed, states
which mediate charged current processes:
\be
W_L = W_1 \, \cos \zeta + W_2 \, \sin \zeta \qquad \qquad W_R = - W_1 \,\sin  \zeta + W_2 \,\cos \zeta ~,
\ee
with the mixing angle $\sin \zeta \sim (v / v_R)^2$ 
proportional to the ratio of the weak scale over the
scale at which the SU(2)$_R$ group is  spontaneously broken,  
$v_R \sim M_{W_2}$, thus leading to the breaking of parity.
To leading order in $(v/v_R)^2$,  this model generates 
the following correlated CC non-standard couplings
\be
\epsilon_L = \epsilon_\mu =  0~, \qquad \epsilon_R = - \zeta~, \qquad  \tilde{\epsilon}_L = - \zeta~, \qquad \tilde{\epsilon}_R = \frac{M_{W_1}^2}{M_{W_2}^2}~, \
\label{eq:lrsm-matching}
\ee
with no dependence on the lepton mass---all other couplings are vanishing.
It is clear then, that this minimal and manifestly left-right symmetric 
model predicts  no appreciable deviations
from the SM in  lepton flavor universality tests. 
On the other hand, the model predicts $\Delta_{\rm CKM} = - 2 \zeta$, so that 
the mixing angle, and therefore the scale of spontaneous parity breaking, 
is strongly constrained by Cabibbo universality tests. Stronger bounds still emerge
from the $K_L$-$K_S$ mass difference~\cite{Beall:1981ze}, though the role of 
long-distance contributions to the mass difference limit the severity of the
constraint~\cite{Donoghue:1983hi,Bigi:1984xh}. 
Given the strong bounds on $\zeta$, the model would also predict unobservably small
effects in decay correlations sensitive to $ |\tilde{\epsilon}_{L,R}|^2 \sim \zeta^2$.

\subsection{\it CC processes in  the MSSM}

Within the MSSM with R-parity, the CC effective couplings $\epsilon_\alpha$
are generated through loop diagrams (vertex corrections and box 
diagrams such as those depicted in Fig.~\ref{fig:box-susy}),
resulting in expressions that are not nearly as simple as the 
ones in Eq.~(\ref{eq:lrsm-matching}).

The chirality flipping couplings $\epsilon_{S,T}$~\cite{Profumo:2006yu} require
the presence of left-right mixing between 
sfermions running in the box diagrams of Fig.~\ref{fig:box-susy},
which is proportional to the small Yukawa couplings or the trilinear soft ``A"  terms.
Ref.~\cite{Profumo:2006yu} analyzed the phenomenological 
constraints on such mixing and determined
 the range of the allowed contributions to the weak decay  coefficients $b$ and $B$,
 arguing that they may provide unique probes of 
left-right mixing in the first generation scalar fermion sector,
provided a precision between $10^{-4}$ and  $10^{-3}$  can be achieved.

Concerning the universality tests,  in  Ref.~\cite{Bauman:2012fx}
it was shown that  the magnitude of 
the corrections  $\Delta_{\rm CKM}$ and $\Delta_{e/\mu}$ can 
be on the order of $10^{-3}$, which is consistent 
with precision electroweak tests and LHC direct searches for supersymmetric particles.
The size of  the universality  violations is controlled by 
the splitting  in the squark versus slepton spectra (Cabibbo universality)
or in the selectron versus smuon spectra (lepton universality).
Moreover,  Ref.~\cite{Bauman:2012fx} showed 
that a comparison of the first row CKM unitarity tests with measurements of
$R_{e/\mu}$ can provide unique probes of the spectrum of 
first generation squarks and first and second generation sleptons,
as illustrated in Figure~\ref{fig:mercedes} and explained in the
figure caption. As a consequence,  universality tests will be powerful diagnostic tools
if supersymmetric partners  are discovered at the LHC.

Finally, a discussion of the 
impact of  Cabibbo and lepton universality tests within the  R-parity violating MSSM
can be found in Refs.~\cite{Kurylov:2001zx,RamseyMusolf:2007yb}.

\begin{figure}[t]
\begin{center}
\vspace{1cm}
\includegraphics[width=0.45\hsize]{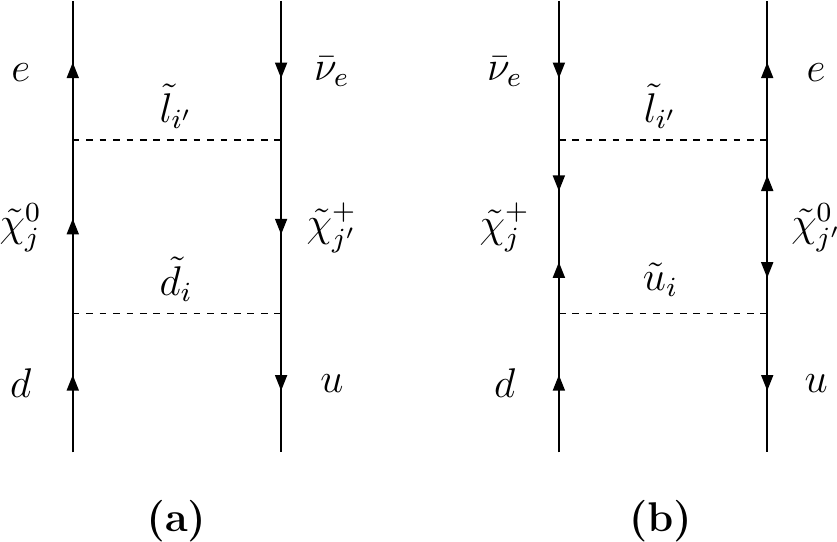}
\begin{minipage}[t]{16.5 cm}
\caption{\small
A subset of the Feynman diagrams giving rise to non-standard CC operators in the MSSM.
\label{fig:box-susy}}
\end{minipage}
\end{center}
\end{figure}

\begin{figure}[t]
\begin{center}
\vspace{1cm}
\includegraphics[width=0.45\hsize]{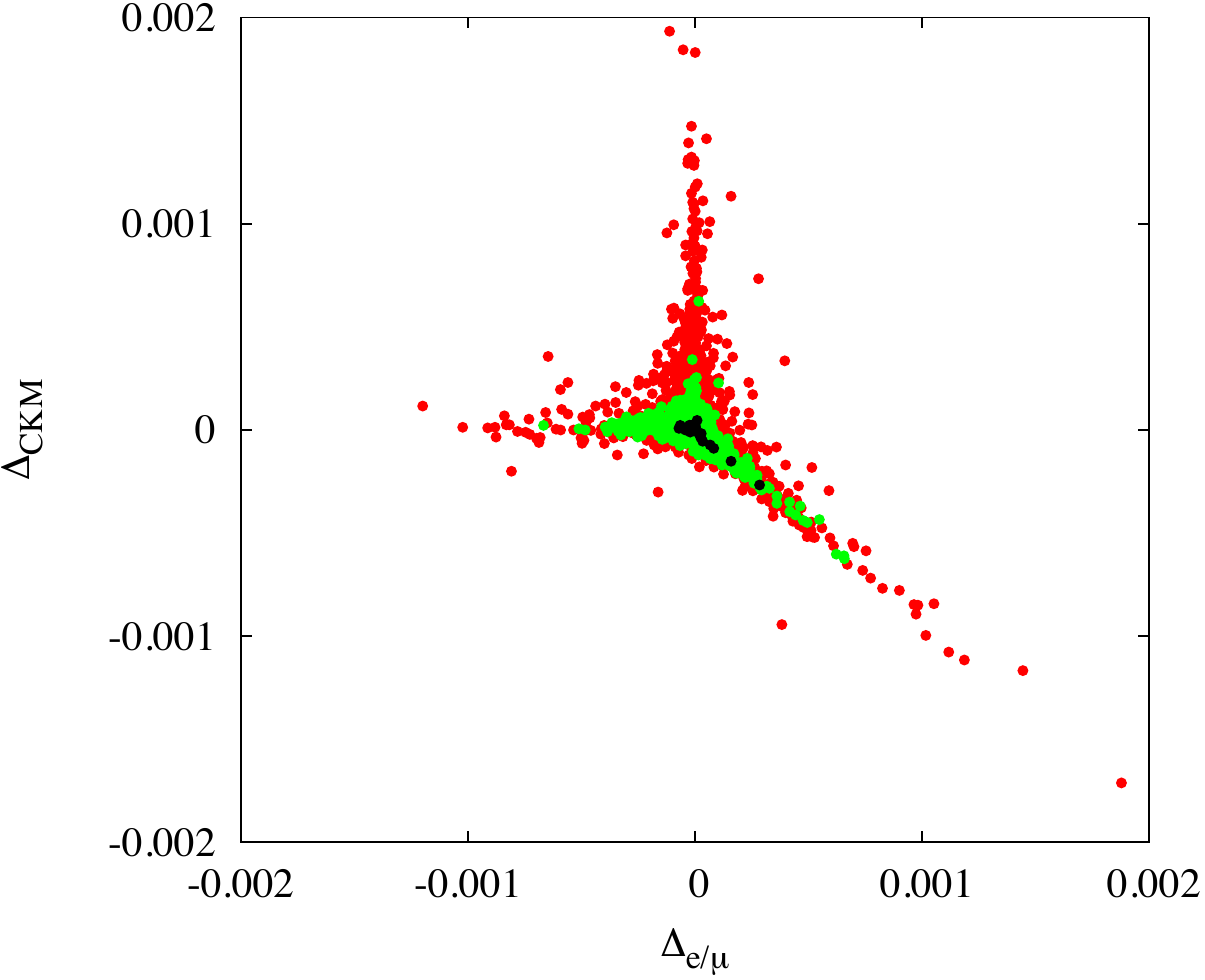}
\begin{minipage}[t]{16.5 cm}
\caption{\small
Correlation between $\Delta_{\rm CKM}$ and $\Delta_{e/\mu}$ in the MSSM.
The red points (dark grey) arise from a generic parameter space scan.
The green points (light grey) 
arise after applying the constraints from  precision electroweak tests.
The black points  arise after applying the constraints from direct searches at the LHC.
The three branches correspond to the following scenarios for the sfermion spectra:
the vertical branch corresponds to light squarks, which 
are been largely ruled out by  the LHC, and heavy sleptons;
the right branch corresponds to light smuons and heavy selectrons and squarks;
the left branch corresponds to light selectrons and heavy smuons and squarks.
Figure reprinted with permission  from  S. Bauman, J. Erler, M. Ramsey-Musolf, ``Charged current universality 
and the MSSM", atXiv:1205.0035 [hep-ph]~\cite{Bauman:2012fx}.
\label{fig:mercedes}}
\end{minipage}
\end{center}
\end{figure}

\section{Conclusions}
\label{sect:conclusions}

In this article we have reviewed the role of precision beta decays measurements
in probing physics beyond the Standard Model 
in the LHC era.
As for all precision tests,  theoretical calculations of the  SM amplitudes
play  a crucial role in setting the stage for BSM searches.
Here we have tried to convey the flavor of the needed theoretical inputs,
emphasizing the increasing role played by lattice QCD both in the 
meson sector, in regard to the determination
of $V_{us}$, 
 and in the nucleon sector, in probing non-standard scalar and tensor couplings through
decay correlations.

Concerning BSM physics, we have emphasized a model-independent EFT approach to beta decays,
assuming that new physics is emergent at high energies, 
based on a  quark-lepton level  effective Lagrangian. 
This approach has a dual benefit.  On one hand, 
it allows an unambiguous comparison of the physics reach of
probes  involving  different hadrons, such as pions and nucleons, and nuclei, limited
only by our ability to compute the requisite matrix elements. 
Moreover, as stated at the start, 
in the absence of an emerging  picture of new dynamics 
from collider searches, 
the EFT analysis is the first necessary step 
to establishing the  motivation and significance of this set of low-energy probes.
The current bounds on the real and imaginary parts of the non-standard couplings 
$\epsilon_{L,R,S,P,T}$ (involving left-handed neutrinos)  and 
$\tilde{\epsilon}_{L,R,S,P,T}$  (involving right-handed neutrinos)
are summarized in Tables~\ref{tab:summaryLRE}, \ref{tab:summaryLIM}, \ref{tab:summaryRRE}, \ref{tab:summaryRIM}. 
The outlook is very positive: 
the effective couplings of 
all the BSM CC operators involving left-handed neutrinos  are currently probed
or will be soon probed  
in low-energy experiments 
at the level of $10^{-3}$ or better. 
This corresponds to probing maximal BSM physics scales 
$\Lambda$ ranging from 7~TeV (for scalar and tensor interactions)
to 11~TeV (for vector interactions),  to ${\cal O}(100)$~TeV (for pseudoscalar interactions,
assuming no cancellations and no mass or Yukawa suppressions).

In all cases, the effective scale probed 
overlaps with the LHC reach: therefore, if new particles are found at the LHC,
beta decays will play an important role in the ``LHC inverse  problem", i.e. in establishing
the properties of the new BSM dynamics. 
This is also explicitly illustrated in the case of the MSSM (see Section~\ref{sect:models}
and Fig.~\ref{fig:mercedes}).
Moreover, if new BSM dynamics is 
above the LHC reach, sometimes termed the ``nightmare scenario,"
one can analyze LHC data  on the process $pp \to e \nu + X$  
in terms of the same EFT used at low energy,
modulo the known QCD running of the various  couplings.
Even in this pessimistic scenario, 
recent theoretical results~\cite{Cirigliano:2012ab}
show that  beta decay measurements
at the $10^{-3}$-level  can be more sensitive than the LHC
in probing non-standard CC interactions, noting once 
again Tables~\ref{tab:summaryLRE}, \ref{tab:summaryLIM}, \ref{tab:summaryRRE}, \ref{tab:summaryRIM}, 
and Fig.~\ref{LHCboundb}, 
and that $10^{-4}$-level measurements would have unmatched sensitivity.
These considerations illustrate 
the relevance of precision beta decay measurements throughout the LHC era.
\\

\noindent \Large{\bf Acknowledgements}}
\vspace{0.3cm}

\noindent 
We thank Michael Ramsey-Musolf for carefully reading the manuscript and  
Mart\'in Gonz\'alez-Alonso and Alejandro Garcia for discussions and correspondence. 
We also thank Doug Bryman, Geoffrey Greene, 
Brad Plaster, and Fred Wietfeldt for comments on the manuscript. 
SG acknowledges partial support from the U.S. Department of Energy under contract DE-FG02-96ER40989.
The work of VC is supported by the U.S. Department of Energy and the LDRD program at Los Alamos National Laboratory.

\bibliographystyle{doiplain}
\bibliography{Cirigliano-BibTex}

\end{document}